\documentclass{aa} 

\usepackage{graphicx}
\usepackage{lipsum}

\usepackage{txfonts}
\usepackage{natbib}
\usepackage{hyperref}
\usepackage{longtable}
\usepackage{supertabular}
\usepackage{placeins}
\usepackage{stfloats}
\usepackage{subcaption}  

\usepackage[normalem]{ulem}

\usepackage{caption}
\usepackage{subcaption}

\usepackage{lscape}

\usepackage[utf8]{inputenc}
\usepackage{booktabs}
\usepackage{hyperref}
\usepackage{multirow}
\usepackage{enumitem}

\usepackage{lscape}

\usepackage[switch]{lineno}

\usepackage{lineno}

\usepackage[dvipsnames]{xcolor}
\definecolor{myblue}{HTML}{1F77B4}
\definecolor{mygreen}{HTML}{2CA02C}
\definecolor{m}{HTML}{D62728}
\definecolor{mymagenta}{HTML}{D33682}
\definecolor{codepurple}{HTML}{C42043}

\newcommand{\program}[1]{\textsc{#1}}

\hypersetup{
unicode=true,           
colorlinks=true,        
citecolor=myblue,       
urlcolor=black        
}
\begin{document}

   \title{Eruptive mass loss less than a year before the explosion of superluminous supernovae}
 \subtitle{II. A systematic search for pre-explosion eruptions with VLT/X-shooter}

    \author{A.~Gkini\inst{1} \href{https://orcid.org/0009-0000-9383-2305}{\includegraphics[scale=0.5]{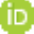}}\and
     C.~Fransson\inst{1} \href{https://orcid.org/0000-0001-8532-3594}{\includegraphics[scale=0.5]{Images/ORCIDiD_icon16x16.pdf}}\and
     R.~Lunnan\inst{1} \href{https://orcid.org/0000-0001-9454-4639}{\includegraphics[scale=0.5]{Images/ORCIDiD_icon16x16.pdf}}\and
    S.~Schulze\inst{2} \href{https://orcid.org/0000-0001-6797-1889}{\includegraphics[scale=0.5]{Images/ORCIDiD_icon16x16.pdf}} \and
   J.~Sollerman\inst{1} \href{https://orcid.org/0000-0003-1546-6615}{\includegraphics[scale=0.5]{Images/ORCIDiD_icon16x16.pdf}}\and 
   K.~Tsalapatas \inst{1} \href{https://orcid.org/0009-0004-1062-8886}{\includegraphics[scale=0.5]{Images/ORCIDiD_icon16x16.pdf}} \and 
   N.~Sarin\inst{3,4} \href{https://orcid.org/0000-0003-2700-1030}{\includegraphics[scale=0.5]{Images/ORCIDiD_icon16x16.pdf}} \and
    M.~Nicholl\inst{5} \href{https://orcid.org/0000-0002-2555-3192}{\includegraphics[scale=0.5]{Images/ORCIDiD_icon16x16.pdf}}\and
    C.~Angus\inst{5} \href{https://orcid.org/0000-0002-4269-7999}{\includegraphics[scale=0.5]{Images/ORCIDiD_icon16x16.pdf}}\and
   U.~Burgaz\inst{6} \href{https://orcid.org/0000-0003-0126-3999}{\includegraphics[scale=0.5]{Images/ORCIDiD_icon16x16.pdf}} \and
    S.~J.~Brennan\inst{7} \href{https://orcid.org/0000-0003-1325-6235}{\includegraphics[scale=0.5]{Images/ORCIDiD_icon16x16.pdf}} \and
   T.-W.~Chen\inst{8} \href{https://orcid.org/0000-0002-1066-6098}{\includegraphics[scale=0.5]{Images/ORCIDiD_icon16x16.pdf}} \and
   A.~Gal-Yam \inst{9} \href{https://orcid.org/0000-0002-3653-5598}{\includegraphics[scale=0.5]{Images/ORCIDiD_icon16x16.pdf}}\and
   A.~Gangopadhyay \inst{1} \href{https://orcid.org/0000-0002-3884-5637}{\includegraphics[scale=0.5]{Images/ORCIDiD_icon16x16.pdf}}\and
   Y.~Hu \inst{1} \href{https://orcid.org/0000-0002-9744-3910}
   {\includegraphics[scale=0.5]{Images/ORCIDiD_icon16x16.pdf}}\and
   M.~M.~Kasliwal \inst{10} \href{https://orcid.org/0000-0002-5619-4938}{\includegraphics[scale=0.5]{Images/ORCIDiD_icon16x16.pdf}}\and
   R.~R.~Laher\inst{11} \href{https://orcid.org/0000-0003-2451-5482}{\includegraphics[scale=0.5]{Images/ORCIDiD_icon16x16.pdf}}\and 
    P.~J.~Pessi\inst{1} \href{https://orcid.org/0000-0002-8041-8559}{\includegraphics[scale=0.5]{Images/ORCIDiD_icon16x16.pdf}}\and 
    B.~Rusholme\inst{11} \href{https://orcid.org/0000-0001-7648-4142}{\includegraphics[scale=0.5]{Images/ORCIDiD_icon16x16.pdf}}\and 
    E.~Russeil \inst{1} \href{https://orcid.org/0000-0001-9923-2407}{\includegraphics[scale=0.5]{Images/ORCIDiD_icon16x16.pdf}}\and 
   A.~Singh \inst{1} \href{https://orcid.org/0000-0003-2091-622X}{\includegraphics[scale=0.5]{Images/ORCIDiD_icon16x16.pdf}}\and
   C.~Skoglund \inst{1} \href{https://orcid.org/0009-0006-7528-0168}{\includegraphics[scale=0.5]{Images/ORCIDiD_icon16x16.pdf}}\and
   R.~Smith \inst{12} \href{https://orcid.org/0000-0001-7062-9726}{\includegraphics[scale=0.5]{Images/ORCIDiD_icon16x16.pdf}}\and
   B.~van Baal \inst{1} \href{https://orcid.org/0009-0001-3767-942X}{\includegraphics[scale=0.5]{Images/ORCIDiD_icon16x16.pdf}}\and
   S.~L.~West \inst{1} \href{https://orcid.org/0000-0002-4575-6157}{\includegraphics[scale=0.5]{Images/ORCIDiD_icon16x16.pdf}}\and
   L.~Yan \inst{10} \href{https://orcid.org/0000-0003-1710-9339}{\includegraphics[scale=0.5]{Images/ORCIDiD_icon16x16.pdf}}
}

   \institute{The Oskar Klein Centre, Department of Astronomy, Stockholm University, Albanova University Center, 106 91 Stockholm, Sweden
   \and
   Center for Interdisciplinary Exploration and Research in Astrophysics (CIERA), Northwestern University, 1800 Sherman Ave, Evanston, IL 60201, USA
  \and
    Kavli Institute for Cosmology, University of Cambridge, Madingley Road, CB3 0HA, UK 
  \and
  Institute of Astronomy, University of Cambridge, Madingley Road, CB3 0HA, UK
  Konkoly Observatory, Research Center for Astronomy and Earth Sciences, H-1121 Budapest Konkoly Th. M. út 15-17., Hungary; MTA Centre of Excellence 
  \and
   Astrophysics Research Centre, School of Mathematics and Physics, Queens University Belfast, Belfast BT7 1NN, UK
   \and
  School of Physics, Trinity College Dublin, College Green, Dublin 2, Ireland 
   \and
  Max Planck Institute for Extraterrestrial Physics, Max-Planck-Gesellschaft, Giessenbachstra{\ss}e 1, Garching, 85748
  \and
  Graduate Institute of Astronomy, National Central University, 300 Jhongda Road, 32001 Jhongli, Taiwan
  \and
    Department of Particle Physics and Astrophysics, Weizmann Institute of Science, 234 Herzl St, 76100 Rehovot, Israel 
    \and
    Division of Physics, Mathematics and Astronomy, California Institute of Technology, Pasadena, CA 91125, USA
  \and
  IPAC, California Institute of Technology, 1200 E. California Blvd, Pasadena, CA 91125, USA 
  \and
  Caltech Optical Observatories, California Institute of Technology, Pasadena, CA 91125, USA 
  }

   \date{}

\abstract
{We present X-shooter spectroscopic and photometric observations of a sample of 21 hydrogen-poor superluminous supernovae (SLSNe-I), spanning a redshift range of $z = 0.13-0.95$, aimed at searching for shells of circumstellar material (CSM). Specifically, we focused on identifying broad \ion{Mg}{II} absorption features that are blueshifted by several thousand kilometers per second relative to the narrow absorption lines associated with the host galaxy. These broad features have previously been interpreted to arise from resonance line scattering of the SLSN continuum by rapidly expanding CSM ejected shortly before explosion. Utilizing high-quality near-ultraviolet spectra, we modeled the region around 2800~\AA\ to characterize the \ion{Mg}{II} line profiles, enabling us to either confirm their presence or place constraints on undetected CSM shells. We identified five objects in our sample that show broad \ion{Mg}{II} absorption features consistent with the presence of CSM. While SN\,2018ibb, SN\,2020xga, and SN\,2022xgc have been previously reported, we identified previously undiscovered CSM shells in DES15S2nr and DES16C3ggu. In the case of DES15S2nr, the CSM shell is located at $\sim 3.4 \times 10^{15}\rm cm$ and is moving with a maximum velocity of $\sim 4800\rm~km~s^{-1}$. For DES16C3ggu, the shell lies at $\sim 4.8 \times 10^{15}\rm cm$ and reaches up to $\sim 4700~\rm km~s^{-1}$. These shells were likely expelled approximately two and three months before the explosion of their respective associated SNe on timescales consistent with late-stage eruptive mass-loss episodes. We further found evidence that the velocities of the CSM shells in all objects lie within $3000-5000~\rm km~s^{-1}$, which may reflect an intrinsic property and could hint at a similar mass-ejection mechanism. We did not find any correlations between the shell properties and the SN properties, except for a marginal correlation between the light curve decline timescale and the shell velocities. This correlation needs further work; however, if it applies, it is a powerful link between the late-time mass ejection and eventual explosion. We further demonstrate that CSM configurations similar to the majority of the detected shells would have been observable in spectra with a signal-to-noise $>5$ per resolution element, and that the lines from a shell are, in general, detectable except in cases where the shell is either very geometrically and/or optically thin. Therefore, we conclude that the non-detections are unlikely to arise from selection effects but they may instead point to the existence of a subclass of SLSN-I progenitors undergoing late-stage shell ejections shortly before explosion.}

\keywords{supernovae: general – supernovae: individual: DES15S2nr, DES16C3ggu, SN\,2018ibb, SN\,2020xga, SN\,2022xgc}

\authorrunning{Gkini et al.}
\titlerunning{Probing eruptive mass-loss prior to superluminous supernovae with VLT/X-shooter}

   \maketitle
%

\section{Introduction}

The evolution and fate of massive stars, including the type of the resulting supernova (SN) explosion, is intimately linked to the mass-loss history during the star’s life. Mass loss in massive stars can begin as early as the main-sequence phase, driven by stellar winds \citep[e.g.,][]{Lucy1970,Castor1975,Lamers1999,Puls2008} with rates depending on factors such as stellar mass \citep[e.g.,][]{Smith2014}, metallicity \citep[e.g.,][]{deJager1988,Vink2001,Hovis2025}, and rotation \citep[e.g.,][]{Maeder2000,Ekstrom2008,Georgy2013,Sibony2024}. In addition to wind-driven mass loss, mass can also be transferred in binary (or higher-order) systems \citep[e.g.,][]{Sana2012,Marchant2024}, where interactions with a companion star can strip substantial amounts of material via Roche-lobe overflow \citep[e.g.,][]{deMink2013,Petrovic2005,Gotberg2017,Laplace2020,Drout2023,Gilkis2025}.

In the late stages of stellar evolution, where evolutionary timescales become much shorter than the timescales for steady mass loss, episodic, violent outbursts can dominate, expelling significant amounts of material. Such eruptive mass-loss events have been observed in extreme cases such as $\eta$ Carinae \citep{Westphal1969}, although the physical mechanisms driving these outbursts remain poorly understood. Several theoretical scenarios have been proposed. One possibility is that hydrodynamic instabilities associated with advanced nuclear burning phases can lead to envelope ejections \citep{Smith2014b}. Another mechanism involves wave-driven mass loss, where convective motions in the C- and O-burning shells excite acoustic waves that propagate outward, depositing energy in the outer layers and driving mass loss \citep{Quataert2012,Shiode2014}. Eruptive mass loss can also result from pulsational pair-instability (PPI; \citealt{Woosley2007,Yoshida2016,Woosley2017,Leung2019,Marchant2019,Renzo2020,Huynh2025}), which occurs in very massive, metal-poor stars with He-core masses between $30$ and $65~\rm M_\odot$. In this scenario, the production of electron–positron pairs in the CO core reduces radiation pressure support, leading to contraction, explosive O burning, and subsequent mass ejection. The strength and the number of pulses increase with the star’s mass, resulting in more massive and energetic ejections \citep{Renzo2020}. 
Eruptive mass-loss episodes can lead to the formation of circumstellar material (CSM) shells around massive stars, composed of material that is expelled at velocities thousands of kilometers per seconds \citep{Renzo2020,Huynh2025}. Following a SN explosion, these shells can become observable either through photometric signatures, such as light curve bumps or undulations \citep[e.g.,][]{Leloudas2012a,Fraser2013,Gomez2019,Hosseinzadeh2022,Chen2023b,West2023}, or spectroscopic features indicative of an interaction with the CSM \citep[e.g.,][]{Ben-Ami2014,Lunnan2016,Yan2017,Pursiainen2022,Aamer2024,Schulze2024,Gkini2024}.

Such shells have been observed in the spectra of superluminous supernovae (SLSNe; \citealt{Quimby2011, GalYam2019a}), a rare class of stellar explosions thought to originate from massive progenitor stars \citep{Moriya2018b,GalYam2012}. These events exhibit peak absolute magnitudes ranging from $-20$ to $-23$~mag \citep{DeCia2018, Lunnan2018b, Chen2023a, Gomez2024}, placing them among the most luminous transients observed. Recently, a distinctive spectroscopic feature has been identified in two SLSNe-I, iPTF16eh \citep{Lunnan2018} and SN\,2018ibb \citep{Schulze2024}, consisting of a second \ion{Mg}{II} absorption line system blueshifted by approximately $3000~\rm~km~s^{-1}$. This feature has been attributed to the resonance scattering of a rapidly expanding CSM shell, which was likely ejected decades prior to the SN explosion. In the case of iPTF16eh, \cite{Lunnan2018} also detected a time- and frequency-dependent \ion{Mg}{II} in emission that moved from $-1600~\rm km~s^{-1}$ to $+2900~\rm km~s^{-1}$ between $100$ and $300$ days after maximum light, and this was attributed to a light echo from that shell. The distance and velocity of the CSM matched well with theoretical predictions of PPI shell ejections. Due to the lack of ultraviolet (UV) spectroscopic data, such a feature has never been observed in typical stripped-envelope SNe, highlighting SLSNe as a powerful diagnostic for studying episodic mass loss in massive stars. However, the detection of these CSM shells has so far been serendipitous, and it is unclear whether this phenomenon is common or characteristic of the broader SLSN population.

To address this question, we conducted a dedicated observational campaign using the X-shooter spectrograph \citep{Vernet2011a} on the ESO Very Large Telescope (VLT) at Paranal Observatory, Chile, with the aim of identifying a
\ion{Mg}{II} absorption line system with properties pointing to shell ejections shortly before the SN explosion.
This paper is the second in a series focused on probing eruptive mass-loss episodes in the progenitors of SLSNe. The first study by \citet{Gkini2025} examined two SLSNe-I, SN\,2020xga and SN\,2022xgc, which were observed as part of the triggered X-shooter program. In both cases, \ion{Mg}{II} absorption features were detected, revealing high-velocity CSM ($\sim 4300$~km~s$^{-1}$) was expelled less than a year prior to explosion. These two detections in \cite{Gkini2025} increased the total number of SLSNe-I exhibiting CSM signatures to four, establishing a growing sample for statistical and theoretical analysis.

For this work, we extended the analysis to include both the non-detections from the X-shooter sample as well as archival spectra of SLSNe-I from the literature, with the aim of constraining the fraction of SLSNe-I that exhibit signatures of eruptive mass loss and of establishing upper limits on the presence of CSM shells. This paper is organized as follows. In Sect.~\ref{sec:obs_data}, we present the SLSN-I sample and the photometric and spectroscopic data analyzed in this study. In Sect.~\ref{sec:spectral_model}, we describe how we modeled the near-UV spectra to search for potential CSM shell signatures hidden within the noise and to place constraints on the shell properties. The connection between the spectroscopic findings, photometry, and overall SN properties is explored in Sect.~\ref{sec:phot_distribution}. We discuss our findings in Sect.~\ref{sec:discussion}, and summarize our results in Sect.~\ref{sec:conclusion}.

Throughout the paper, the photometric measurements are reported in the AB system and the uncertainties are provided at 1$\sigma$ confidence. We assume a flat Lambda cold dark matter cosmology with $H_{0} = 67.4$~km~s$^{-1}$~Mpc$^{-1}$, $\Omega_{m} = 0.31$, and $\Omega_{\Lambda} = 0.69$ \citep{Planck2020}.

\section{The X-shooter sample} \label{sec:obs_data}

\setlength{\tabcolsep}{1.2pt}
\begin{table*} [ht!]
\centering
\caption{SLSN-I X-shooter sample.}\label{tab:sample}
\begin{tabular*}{1\linewidth}[!ht]{@{\extracolsep{\fill}}lcccclll}
\hline
\hline
Name & R.A. & Decl. & $z$ & $E(B-V)_{MW}$ & Discovery\tablefootmark{\scriptsize a} & Classification & Reference\\
 & (J2000.0) & (J2000.0) & & (mag) & Group & & \\
\hline
SN\,2020rmv & 00:40:00.19 & -14:35:25.03 & $0.26$ & $0.019$ & ATLAS & \cite{Terrenan2020a} & \cite{Chen2023a} \\
SN\,2020xga & 03:46:39.37 & -11:14:33.90 & $0.43$ & $0.049$ & PS1 & \cite{Gromadzki2020a} & \cite{Gkini2025} \\
SN\,2020zbf & 01:58:01.67 & -41:20:51.84 & $0.20$ & $0.014$ & ATLAS & \cite{Ihanec2020a} & \cite{Gkini2024}  \\
SN\,2020abjx & 02:15:02.31 & -08:37:43.61 & $0.39$ & $0.022$ & ZTF & \cite{Yan2020a} & \cite{Gomez2024}\\
SN\,2020abjc & 09:28:00.28 & +14:07:16.60 &  $0.22$ & $0.028$  &  ZTF & \cite{Blanchard2020a} & \cite{Gomez2024}\\
SN\,2021ek & 03:23:49.90 & -10:02:41.44 &  $0.19$ & $0.053$ &  ZTF &  \cite{Gillanders2021} & \cite{Gomez2024}\\
SN\,2021fao & 10:28:42.55 & -11:02:34.27 &  $0.28$ & $0.048$  & ATLAS & \cite{Gkini2025b} & This paper\\
SN\,2021hpx & 09:31:06.26  & -19:31:04.86  & $0.21$  & $0.046$ & ATLAS & \cite{Gonzalez2021a} & \cite{Gomez2024}  \\
SN\,2022abdu &  03:06:04.89 & -46:43:16.90 & $0.13$ & $0.012$ & ATLAS & \cite{Gromadzki2022a} & \cite{Gomez2024}  \\
SN\,2022acch & 09:24:01.91 
 & -13:31:29.06 & $0.42$ & $0.040$ & ZTF & \cite{Gkini2022} &This paper  \\
SN\,2022xgc & 07:12:41.81 & +07:18:59.95 & $0.31$ & $0.061$ & ZTF & \cite{Gromadzki2022a} & \cite{Gkini2025} \\

\hline
\hline
LSQ12dlf &  01:50:29.80 & -21:48:45.40 & $0.26$ & $0.011$ & LSQ & \cite{Inserra2012a} & \cite{Nicholl2014}\\
SN\,2013dg & 13:18:41.38  & +07:04:43.10 & $0.26$ & $0.048$ & CSS & \cite{Smartt2013} & \cite{Nicholl2014}\\
iPTF13ajg & 16:39:03.95 & +37:01:38.40 & $0.74$ & $0.012$  & iPTF & \cite{Vreeswijk2014} &\cite{Vreeswijk2014}  \\
iPTF15cyk & 07:42:14.87 & +20:36:43.40 & $0.54$ & $0.051$ & iPTF & \cite{Kasliwal2016} & \cite{Kasliwal2016}\\
OGLE15qz & 03:08:35.88  & -70:30:41.7  & $0.59$ & $0.028$ & OGLE & \cite{Kostrzewa2015} & Aryan et al., in prep.\\
DES15S2nr & 02:40:44.62 & -00:53:26.40  & $0.22$ & $0.030$ & DES & \cite{Angus2019} & \cite{Angus2019}\\
DES16C3dmp & 03:31:28.35  & -28:32:28.30 & $0.57$ & $0.007$ & DES & \cite{Angus2019} & \cite{Angus2019}\\
DES16C3ggu & 03:31:12.00 & -28:34:38.70 & $0.95$ & $0.007$ & DES & \cite{Angus2019} & \cite{Angus2019}\\
SN\,2018ibb & 04:38:56.95  & -20:39:44.10 & $0.16$ & $0.030$ & ATLAS & \cite{Pursiainen2018a} & \cite{Schulze2024}\\
SN\,2021gch & 10:27:25.276  & +20:27:15.92 & $0.51$ & $0.018$ & ZTF & \cite{Lunnan2024} & This paper\\

\hline
\end{tabular*}
\tablefoot{The upper half of the table lists all the objects triggered through our X-shooter programs, while objects in the lower half are from the literature and ESO archive. \tablefoottext{a}{The LSQ stands for La Silla-QUEST survey, CSS for Catalina Sky Survey, and OGLE for Optical Gravitational Lensing Experiment.} }
\end{table*}

Motivated by the discovery of CSM shells in iPTF16eh \citep{Lunnan2018} and SN\,2018ibb \citep{Schulze2024}, we collected a spectroscopic sample of 19 objects with the medium-resolution ($R \sim 5400$) X-shooter spectrograph ($\rm program~IDs$: $\rm 105.20PN$, $\rm 106.21L3$, $108.2262$ and $\rm 110.247C$), as the spectral resolution of this instrument is essential for distinguishing between the interstellar medium (ISM) and CSM absorption features. The selection criteria for the program focused on objects that:
\begin{enumerate}
    \item had been spectroscopically classified as SLSNe-I from other facilities,
    \item were observable from Paranal,
    \item had a redshift $z>0.11$ to ensure that the \ion{Mg}{II} resonance lines were reachable with X-shooter, and
   \item were brighter than a magnitude cut-off ($18-19.5$~mag) to ensure a S/N > 10. 
\end{enumerate}
We note that the last criterion was not satisfied for all objects due to weather conditions, and many of the spectra in our sample have S/N < 10.

Six objects from our initial triggered sample of 19 were excluded from the analysis. Among them, two objects, SN\,2021aaev \citep{Hu2025} and SN\,2021adxl \citep{Brennan2024}, were classified as SLSNe-II and are therefore not included in the present H-poor SLSN sample. One event, SN\,2021yfj, was identified as the first Type Ien SN, almost reaching the luminosity of a SLSNe \citep{Schulze2025}. In addition, SN\,2022czy \citep{Blanchard2022} and SN\,2022csn \citep{Arcavi2022} were subsequently reclassified as tidal disruption events, while SN\,2023ayq \citep{Schulze2023a} was reclassified as a SN Ia-CSM. From the remaining triggered sample of 13, we identified 11 SLSNe-I in which \ion{Mg}{II} absorption features from the CSM are not clearly detected (see Sect.~\ref{sec:model_nondete}), and two events where such absorption systems from the CSM shell are observed and analyzed in detail by \cite{Gkini2025}. 

Out of the 46 SLSNe-I that occurred during the observational window of our X-shooter program (October 2020–March 2023), 26 were accessible from Paranal at an airmass below 2 and had redshifts $z > 0.11$. However, due to a number of selection constraints, including technical factors such as program availability in a given semester, interruptions caused by the COVID-19 shutdown, sources becoming too faint by the time of classification, and the requirement for prior classification by other facilities, we ultimately triggered 13 SLSNe-I. In conclusion, while the final triggered sample of 13 objects is not complete, it still provides a robust dataset for investigating the presence of CSM in SLSNe-I.

The sample is supplemented with SLSNe-I that have high-quality X-shooter spectra available in the literature and ESO archive\footnote{Program ID: 105.20CB.002, PI: Annalisa De Cia}. To identify these objects, we conducted a thorough review of all publicly available SLSN-I spectra in the WISeREP and ESO archive\footnote{https://www.wiserep.org/} \citep{Yaron2012a} up to October 2024. Although the first SLSN-I discovered to exhibit a CSM shell was observed with the LRIS instrument mounted on the Keck I telescope, the \ion{Mg}{II} absorption lines associated with the CSM were unresolved in that dataset. Thus, we restrict our analysis to X-shooter spectra. Out of a total of 281 known SLSNe-I (reported in TNS), we identified 20 objects with publicly available X-shooter spectra, 19 of which fulfill our redshift limit.

A signal-to-noise (S/N)-based quality cut was applied to the full sample of 32 (13 + 19) SLSNe-I. For each spectrum, the S/N was computed as the median of the flux divided by the associated error at each wavelength within the $2675-2875$~\AA\ interval. We excluded all spectra with S/N $< 3$ per resolution element at the expected location of Mg II, as spectra below this threshold do not yield reliable or informative results regarding the presence or absence of a CSM. After implementing the quality cut criteria, the final sample comprises 11 SLSNe-I from our initial triggered observations (SN\,2021txk and SN\,2021oes are excluded due to S/N of 1.2 and 1.5 per resolution element, respectively) and 10 SLSNe-I from the literature, resulting in a total of 21 objects. The final sample is presented in Table~\ref{tab:sample}.

\subsection{Spectroscopy}

\begin{figure*}[!h]
    \centering
\includegraphics[width=1\textwidth]{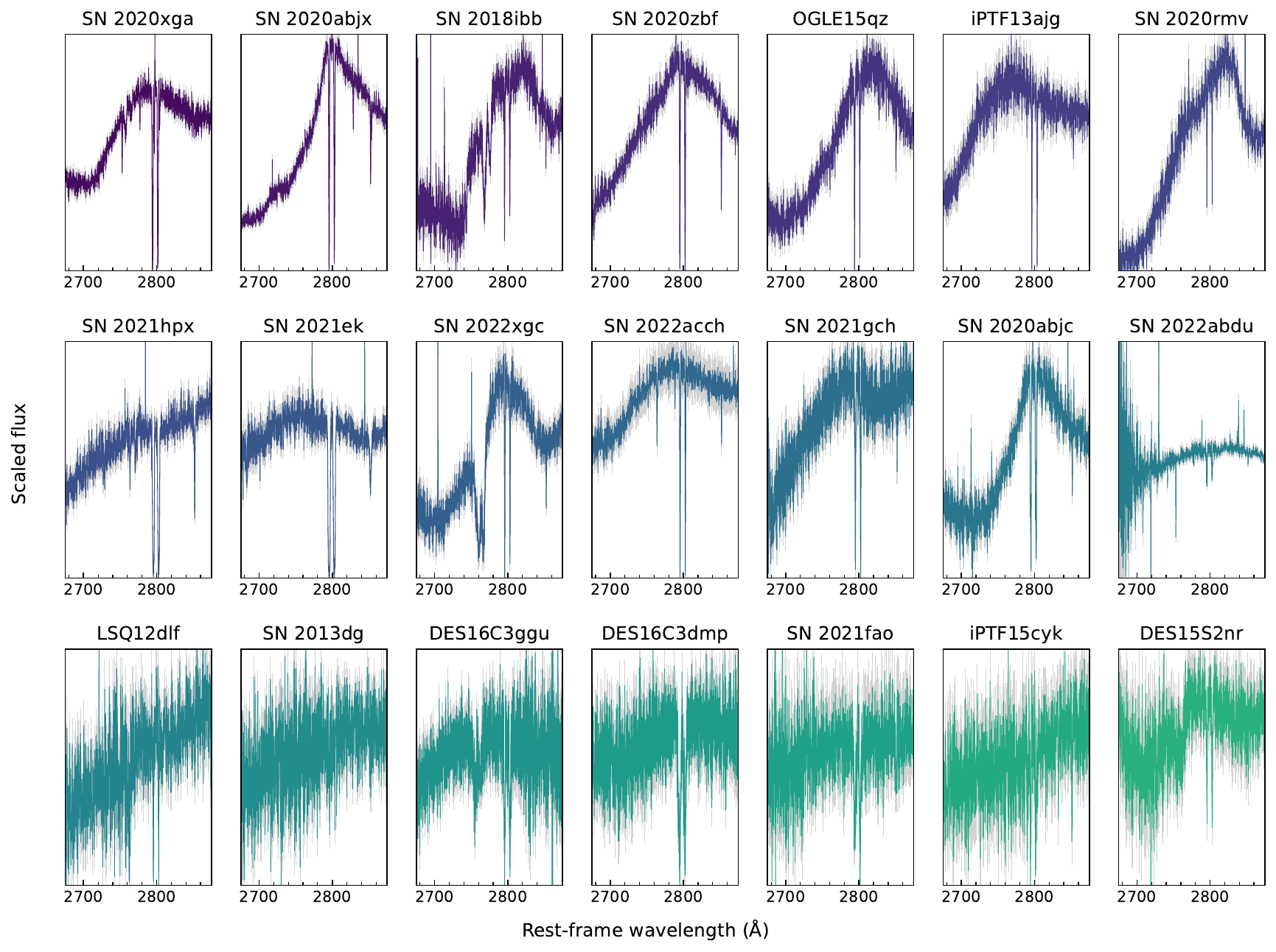}

    \caption{Near-UV spectroscopic observations of the 21 SLSNe-I in the final X-shooter sample analyzed in this study. Each panel shows the observed spectrum of a single object around the 2800~\AA\ region, with the corresponding $1\sigma$ uncertainty shaded in gray. All spectra have been corrected for Milky Way extinction. The objects are presented in order of descending S/N.}
    \label{fig:spectra}
\end{figure*}

All the X-shooter observations were carried out using ultraviolet (UVB), visible (VIS), and near-infrared (NIR) arms using 1\farcs0, 0\farcs9, and 0\farcs9 wide slits, respectively. The observations were conducted in nodding, stare, or offset mode, covering a wavelength range from 3000 to 24000~\AA. Since the all data were reduced in different ways, we retrieved the raw data of all objects from the ESO archive and performed our own data reduction. The reduction process began with the removal of cosmic rays using the \texttt{astroscrappy}\footnote{\url{https://github.com/astropy/astroscrappy}} package, which implements the algorithm of \citet{vanDokkum2001a}. Next, the data were reduced with version 3.6.3 of the X-shooter pipeline using the ESOReflex workflow engine \citep{Goldoni2006a, Modigliani2010}. For the objects that were observed in the nodding mode, the UVB and VIS-arm data were reduced in stare mode to boost the S/N within the wavelength range relevant to our analysis. This step proved to be essential, as an improved S/N significantly increases sensitivity to weak spectral features, which is critical for detecting the presence of a CSM shell (see Sect.~\ref{sec:spectral_model}). The corrected two-dimensional spectra were then co-added using the reduction tools developed by \citet{Selsing2019a}\footnote{\url{https://github.com/jselsing/XSGRB_reduction_scripts}}. To achieve proper skyline subtraction, the NIR-arm data were processed in nodding mode. The wavelength calibration of all spectra was adjusted to account for barycentric motion. The spectra from the different arms were combined by averaging the overlapping regions. Each spectrum was flux calibrated against standard stars. All spectra used in this study were corrected for Milky Way extinction and are absolutely calibrated using contemporaneous photometric measurements. The spectra are presented in Fig.~\ref{fig:spectra}. A log of the spectroscopic observations is provided in Table~\ref{tab:spectra_log}, and the fully reduced spectra have been uploaded to the WISeREP archive.

\subsection{Photometry}

Photometric measurements of our triggered X-shooter sample were conducted with the Zwicky Transient Facility (ZTF; \citealt{Bellm2019,Graham2019,Dekany2020}) survey. The ZTF forced point spread function-fit photometry was retrieved from the Infrared Processing and Analysis Center (IPAC; \citealt{Masci2019a}) for the $gri$ bands. To construct rest-frame light curves, we followed the ZTF data processing guidelines\footnote{\url{https://irsa.ipac.caltech.edu/data/ZTF/docs/ZTF_zfps_userguide.pdf}}, which include baseline correction, validation of flux uncertainties, nightly co-addition of measurements, and conversion of differential fluxes to the AB magnitude system. A 3$\sigma$ quality cut was applied to the photometric data. We note that among our triggered sample, only SN\,2020rmv is included in the ZTF sample presented by \citet{Chen2023a} due to the time threshold in October 2020.

For the objects in our triggered sample where ZTF photometry was not available, we retrieved forced photometry from the ATLAS forced photometry server\footnote{\url{https://fallingstar-data.com/forcedphot/}} \citep{Tonry2018a,Smith2020a,Shingles2021a} for both $c$ and $o$ filters. The clipping and binning, with a bin size of 1 day, of the ATLAS data were done using the \texttt{plot\_atlas\_fp.py}\footnote{\url{https://gist.github.com/thespacedoctor/86777fa5a9567b7939e8d84fd8cf6a76}} python script. Similarly to the ZTF data, we removed the measurements with $< 3\sigma$ significance and converted the resulting fluxes to the AB magnitude system. 

For the ten supplemented objects, we used the photometric data reported in the original studies referenced in Table~\ref{tab:sample}, with the exception of iPTF15cyk and SN\,2021gch. For iPTF15cyk, we retrieved the light curves in the $BRgiz$ filters from IPAC \citep{Laher2014}. The data were reduced using the \program{PTFID} pipeline\footnote{\url{https://web.ipac.caltech.edu/staff/fmasci/ptf/masci_iPTFworkshop2014_ptfide.pdf}} for the P48 instrument and the reduction pipeline described by \citet{Fleming2016} for the P60 instrument. For SN\,2021gch we retrieved the ZTF data from IPAC in the $gri$ bands and we followed the same procedure as for our triggered X-Shooter sample.

\section{Spectral modeling} \label{sec:spectral_model}

To date, four SLSNe-I have been reported in the literature \citep{Lunnan2018, Schulze2024, Gkini2025} that exhibit 
a \ion{Mg}{II} absorption line system that move slower than the SN ejecta ($\sim 4000$ vs $\sim 10000~\rm km~ s^{-1}$) but has broader lines than those originating from gas in the host ISM  ($250$–$500~\rm km~s^{-1}$ vs <$100~\rm km~s^{-1}$; \citealt{Kruhler2015, Arabsalmani2018}). These characteristics suggest that the blueshifted features arise from fast-moving material that is physically distinct from both the SN ejecta and the host ISM. For a more detailed discussion of these features, see \citet{Lunnan2018} and \citet{Gkini2025}.

Modeling of the CSM \ion{Mg}{II} lines in these four objects revealed that the CSM shells exhibit a broad range of properties: velocities $v_{\rm max}$ between $3300$ and $4400~\rm km~s^{-1}$, inner radii $R_{\rm in}$ spanning from $1.8$ to $48.1~R_{\rm ph}$ where $\rm R_{\rm ph}$ is the photospheric radius of the SN, outer radii from $2.0$ to $50.7~R_{\rm ph}$, and optical depths $\tau$ ranging from $0.5$ to $10$. These variations in shell parameters are reflected in the diversity of the line profiles (see \citealt{Gkini2025}, their Fig.~14). In the specific case of SN\,2020xga \citep{Gkini2025}, the shell was only marginally detected, prompting a discussion of the necessary physical conditions for a CSM shell to be discernible above the noise level in the SN spectrum. This consideration, along with the absence of such a \ion{Mg}{II} absorption system in the majority of the observed SLSNe-I, motivated a systematic investigation of the detectability limits of CSM shells in SN spectra.

\subsection{Monte-Carlo scattering code} \label{sec:montecarlo_code}

The modeling of the CSM \ion{Mg}{II} lines in SN\,2018ibb, SN\,2020xga, and SN\,2022xgc was performed using the Monte Carlo scattering code described by \citet{Gkini2025}, that is based on \cite{Fransson2014a} and \cite{Taddia2020}. As mentioned in \cite{Gkini2025}, the code assumes homologous expansion of a shell with velocity $v= v_{\rm max}(r/R_{\rm out})$, where $v_{\rm max}$ denotes the maximum shell velocity and $R_{\rm out}$ is the shell’s outer radius. This velocity structure is appropriate for modeling material expelled in a short-duration, eruptive mass-loss episode. This assumption would break down if the SN ejecta was interacting with the shell, but in the absence of any observational signatures of such interaction, in both the spectra and multiwavelength photometry, the homologous assumption remains the most self‑consistent choice. Indeed, the \ion{Mg}{II} absorption profiles in all SLSNe with putative CSM were best modeled under the homologous hypothesis. In this regime, the blue and red velocity edges of the shell absorption, $v_{\rm blue}$ and $v_{\rm red}$ respectively, map directly onto the shell’s inner and outer radii as follows:
\begin{equation*}
\begin{split}
    v_{\rm blue} &  = -v_{\rm max}~\rm 
    \\
    v_{\rm red} &  = -v_{\rm max} \left(\frac{R_{\rm in}}{R_{\rm out}}\right) \left[1 - \left(\frac{R_{\rm ph}}{R_{\rm in}}\right)^2\right]^{1/2}.
    \\
\end{split}   
\label{eq:vel_br}
\end{equation*}
Thus, determining the blue and red velocities of the absorption we can estimate the radius $R_{\rm in}$ and $R_{\rm out}$ of the shell relative to the photospheric radius:

\begin{equation*}
    \frac{R_{\rm in}}{R_{\rm out}} =   
    \left[ \left(\frac{v_{\rm red}}{v_{\rm blue}}\right)^2 + \left(\frac{R_{\rm ph}}{R_{\rm out}}\right)^2 \right]^{1/2}.
\label{eq:vel_br}
\end{equation*}

We modeled the CSM shell with a uniform optical depth $\tau$, which under the Sobolev approximation \citep{Sobolev1957} is given by
\begin{equation}
    \tau=\frac{g_2 \lambda^3 A_{21} n_{1} t}{8 \pi g_1 } .
    \label{eq:tau}
\end{equation}
Here, $A_{21} = 2.8 \times 10^8$ s$^{-1}$ is the transition rate between the upper level 2 and lower level 1, $n_{1}$ is the number density in the ground state, $\lambda$ the wavelength, $g_1=2$ and $g_2$ the statistical weights of the lower and upper levels, respectively, and $t$ the time from the eruption. Given the shells’ narrow geometry, plausible density gradients such as a wind profile ($\rho \propto r^{-2}$) would only marginally affect the absorption signatures. Although this uniform $\tau$ assumption is a simplification, it is a reasonable choice given the significant uncertainties in the progenitor’s mass-loss history. Alternative density profiles or nonspherical geometries could certainly be explored, but doing so is beyond the scope of this study. It would require dedicated modeling of different mass‑loss prescriptions, the ionization and the temperature structure. Moreover, the well‑characterized, spherically symmetric shell inferred for iPTF16eh \citep{Lunnan2018} motivates our continued use of a spherical geometry in this analysis.

Finally, using $\Delta r = (v_{\rm max}-v_{\rm in})~t$ for a homologous shell we can estimate the column density of \ion{Mg}{II}, $N=n_1 \Delta r$, 
\begin{equation}
    N=2.2 \times 10^{14} \left(\frac{v_{\rm max}}{10^3 ~\rm km~s^{-1}}\right) \left(1 - \frac{R_{\rm in}}{R_{\rm out}}\right) \tau \  \ \rm cm^{-2}.
\label{eq:density}
\end{equation}  
In the highly optically thick regime, this equation no longer provides an absolute column density measurement but instead yields only a lower limit.

In this study, while the core functionality of the code remains unchanged, it has been adapted for this work to allow exploration of a significantly broader parameter space. The code requires as input a set of parameters: $v_{\rm max}$, $R_{\rm in}$, $R_{\rm out}$, and $\tau$ along with a prescribed ``continuum'' spectrum, and outputs the resulting \ion{Mg}{II} line profile emerging from a spherically symmetric CSM shell defined by the input parameters. The ``continuum'' represents photons produced near 2800~\AA\ in the SN photosphere that are scattered by the expanding shell. Whereas in \citet{Gkini2025} the background continuum was shaped by the broad \ion{Mg}{II} features originating in the SN ejecta, in this work we assume a flat continuum. This approach enable us to run a large ($\sim 200\,000$ models) parameter grid reducing the computational expense. Although this simplification neglects the evolution of temperature and velocity of the ejecta, which primarily shape the overall continuum itself, it has no impact on the CSM \ion{Mg}{II} P‑Cygni absorption profiles. Since each synthetic absorption profile is ultimately multiplied by the object‑specific continuum (see Sect.~\ref{sec:model_nondete}), the modeling assuming a flat continuum does not affect our ability to detect the CSM signatures. 

\subsection{Modeling of the full X-Shooter sample} \label{sec:model_nondete}

\subsubsection{Grid of parameter space}

\begin{figure*}[!ht]
  \centering

  \begin{subfigure}[b]{0.5\textwidth}
    \centering
    \includegraphics[width=\textwidth]{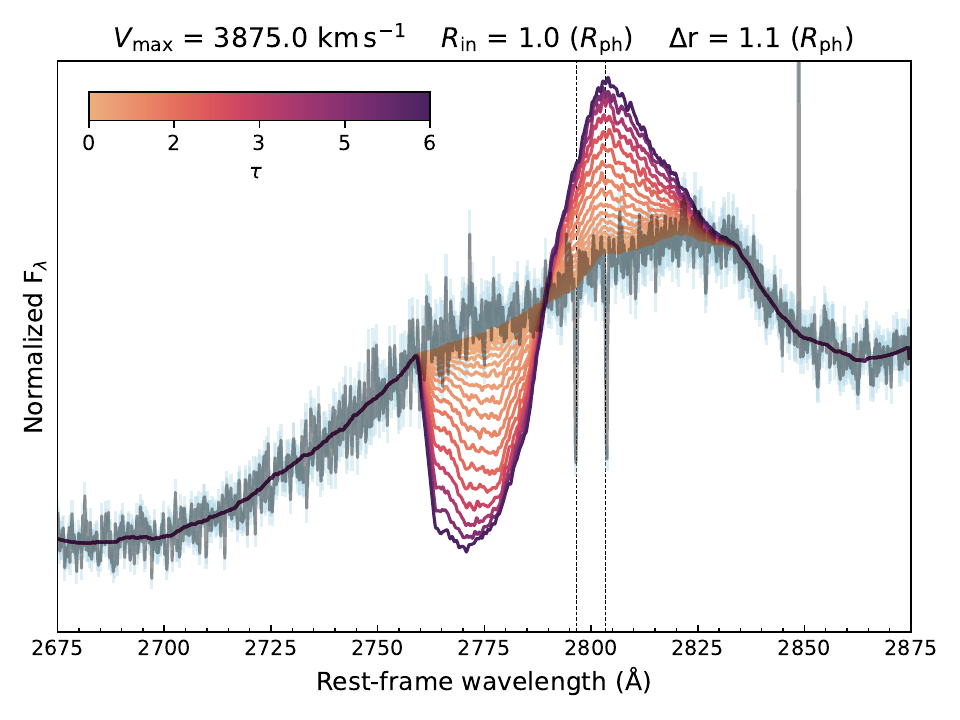}
    \caption{}
    \label{fig:parameter_grid_tau}
  \end{subfigure}
  \hspace{-0.8em} 
  \begin{subfigure}[b]{0.5\textwidth}
    \centering
    \includegraphics[width=\textwidth]{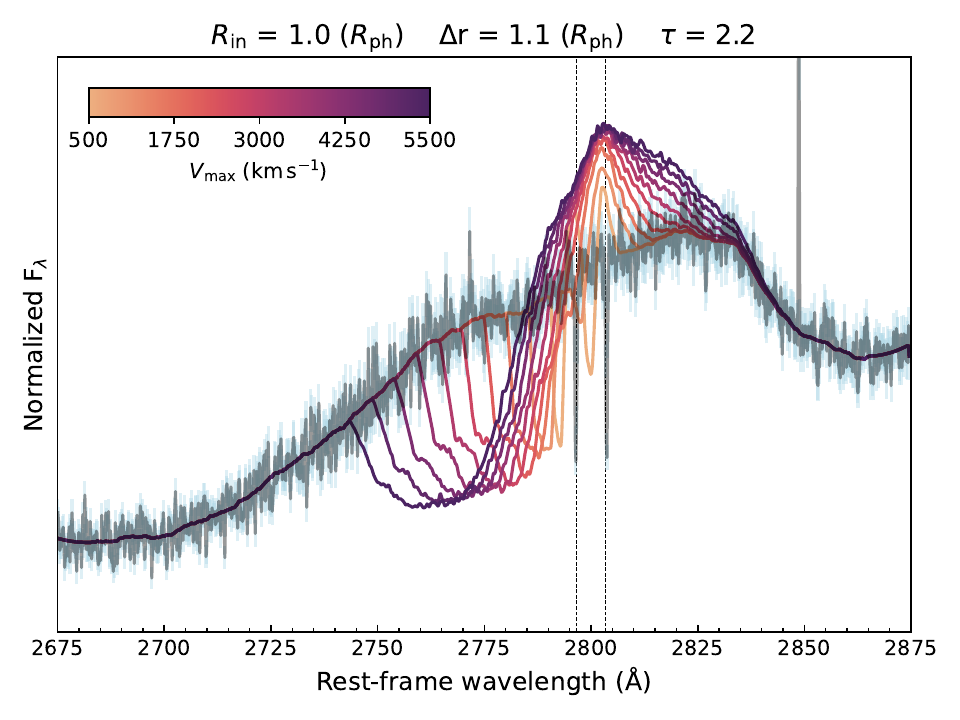}
    \caption{}
    \label{fig:parameter_grid_vel}
  \end{subfigure}
  \hspace{-0.8em} 
  \begin{subfigure}[b]{0.5\textwidth}
    \centering
    \includegraphics[width=\textwidth]{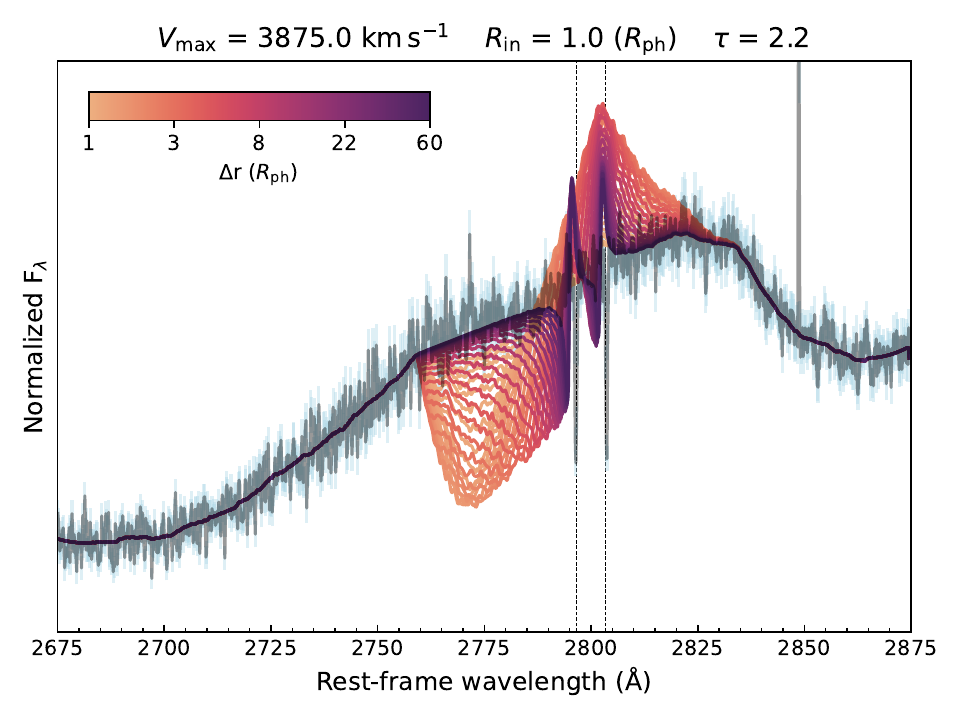}
    \caption{}
    \label{fig:parameter_grid_rout}
  \end{subfigure}
  \hspace{-0.8em}
  \begin{subfigure}[b]{0.5\textwidth}
    \centering
    \includegraphics[width=\textwidth]{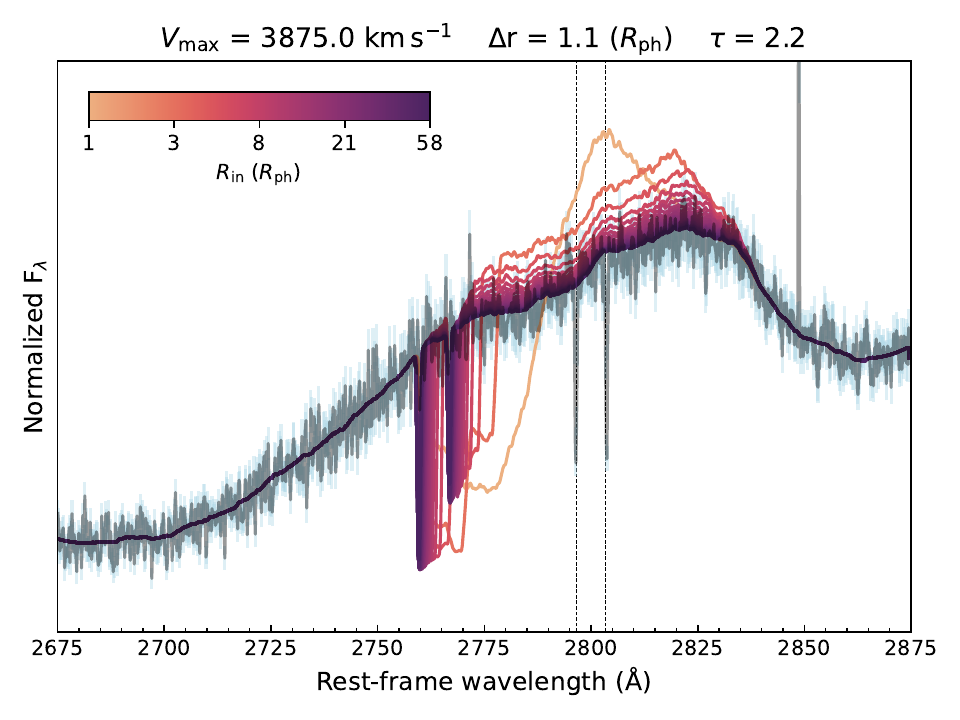}
    \caption{}
    \label{fig:parameter_grid_rin}
  \end{subfigure}

  \caption{Synthetic spectra generated using the Monte Carlo scattering code, where three of the four parameters were held fixed while the fourth was varied: optical depth $\tau$ in panel (a), maximum velocity $v_{\rm max}$ in panel (b), shell width $\Delta r$ in panel (c), and inner radius $R_{\rm in}$ in panel (d). The colorbar indicates the value of the parameter being varied in each case, while the values of the fixed parameters are noted in the title of each panel. The $\Delta r$ and $R_{\rm in}$ are in units of photospheric radius $R_{\rm ph}$ and the colorbars are in logarithmic scale for better visualization. For clarity, the host galaxy \ion{Mg}{II} lines (black dashed lines) have been excluded from the model spectra. The gray curve shows the observed spectrum of SN\,2020rmv, with the light blue shaded region indicating the 1$\sigma$ uncertainty.}
  \label{fig:parameter_grid}
\end{figure*}
To investigate the properties of CSM shells that may be hidden in the noise in the spectral region around $2800$~\AA\ in our sample, we conducted a parameter survey exploring a range of shell configurations. Specifically, we varied $v_{\rm max}$, $R_{\rm in}$, thickness ($\Delta r~(R_{\rm ph})$), and $\tau$. To systematically explore this parameter space, we constructed a grid defined by:

\begin{equation*}
\begin{split}
    v_{\rm max}~(\rm km~s^{-1}) &  \in (500,5500)
    \\
    R_{\rm in}~(R_{\rm ph}) &  \in (1, 60)
    \\
    \Delta r~(R_{\rm ph}) &  \in (0.04, 59)
    \\
    \tau &  \in  (0.05, 6)
\end{split}  
\end{equation*}
We set the lower limit of $v_{\rm max}$ to $500~\rm km~s^{-1}$ to avoid overlap between potential CSM features and the host galaxy's \ion{Mg}{II} absorption system. Velocities above $5500~\rm km~s^{-1}$ were excluded, as they begin to approach SN ejecta velocities.  For $R_{\rm in}$, we explored a range starting from shells attached to the SN photosphere, extending up to $60~R_{\rm ph}$ to also include cases such as iPTF16eh, where the shell was located at approximately $50~R_{\rm ph}$. Rather than setting the outer radius directly, we define the shell thickness $\Delta r$, so that $R_{\rm out} = R_{\rm in} + \Delta r$. This setup enables the exploration of a wider range of shell sizes.  For $\tau$ we imposed an upper limit of $6$, beyond which the results show negligible change. To optimize computational efficiency in this parameter grid exploration, we sampled both $\Delta r$ and the $\tau$ logarithmically. This logarithmic sampling allows a more thorough investigation of the effects of optically and/or geometrically thin shells.

To understand how each of the modeled parameters, $v_{\rm max}$, $R_{\rm in}$, $\Delta r$, and $\tau$, influences the synthetic spectra and visualize these effects, we present a series of example synthetic spectra for SN\,2020rmv in Fig.~\ref{fig:parameter_grid}, varying one parameter at a time while keeping the others fixed. As shown in Fig.~\ref{fig:parameter_grid_tau}, increasing the optical depth $\tau$ (at fixed $v_{\rm max}$, $R_{\rm in}$, and $\Delta r$) enhances the depth and the prominence of the absorption features. At the same time, the shell begins to contribute significantly to the emission, modifying the shape of the underlying continuum. In contrast, shells with low optical depths have a negligible impact on the spectrum and remain largely consistent with the SN continuum, since the contribution from the shell is minimal. In Fig.~\ref{fig:parameter_grid_vel}, varying the velocity of the CSM shell shifts the absorption features to higher velocities and broadens the lines. This effect is due to the definition of the homologous expansion, expressed as $v_{\rm max} - v_{\rm in} = v_{\rm max}[1 - (R_{\rm in}/R_{\rm out})]$. When the ratio $R_{\rm in}/R_{\rm out}$ is held constant, increasing $v_{\rm max}$ leads to broader absorption features, and as a result the doublet is blended.

Varying the shell width $\Delta r$ also has a significant impact on the shape of the synthetic \ion{Mg}{II} CSM lines as shown in Fig.~\ref{fig:parameter_grid_rout}. Broader shells, extending well beyond the photospheric radius, span a wider range of velocities (from $-v_{\rm max}$ to nearly zero). In these cases, the absorption features appear sharper, with the maximum absorption primarily originating from the slower-moving inner regions of the shell. The associated emission component becomes narrower, as most photons are scattered out of the line of sight. In contrast, shells with widths comparable to the photospheric radius sample a more limited velocity range (depending on the $R_{\rm out}$) and produce wider absorption features.  In such narrow shells, the \ion{Mg}{II} doublet often blends into a P Cygni-like profile, resembling those commonly observed in expanding SN ejecta. The line profiles are also sensitive to variations in the inner shell radius $R_{\rm in}$ when the other parameters are held constant (see Fig.~\ref{fig:parameter_grid_rin}). As $R_{\rm in}$ increases, placing the shell farther from the photosphere, the absorption features become narrower.  This is because the shell's inner edge moves faster (closer to $v_{\rm max}$) due to its greater distance while $v_{\rm max}$ remains constant.  At the same time, the shell contributes little to the emission component, resulting in a flatter profile that remains consistent with the continuum level.

\subsubsection{Observational constraints on CSM properties} \label{sec:rule_out}

\begin{figure*}
  \centering
    \subcaptionbox{SN\,2021fao}{\includegraphics[width=0.45\textwidth]{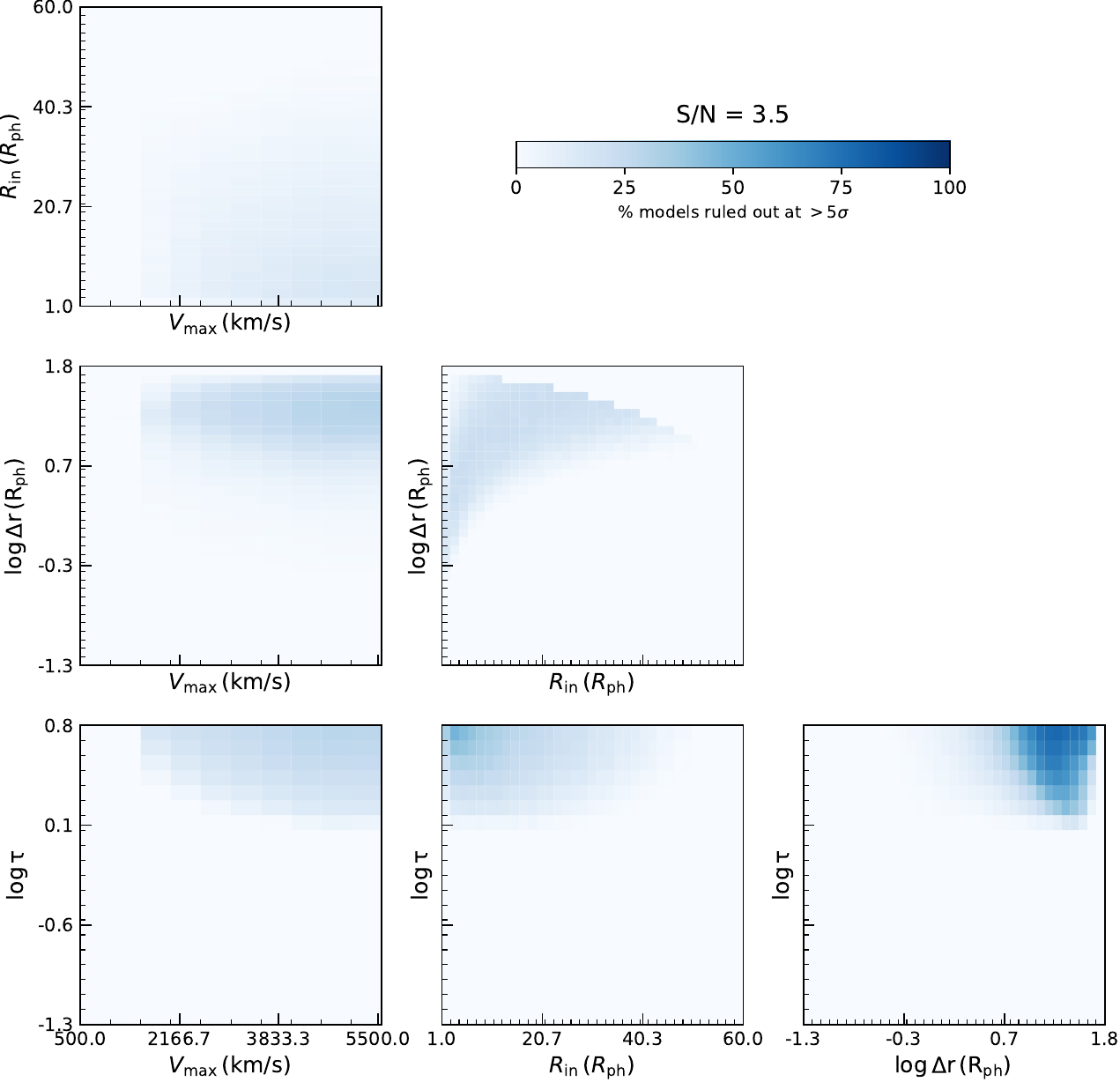}}
    \subcaptionbox{SN\,2022abdu}{\includegraphics[width=0.45\textwidth]{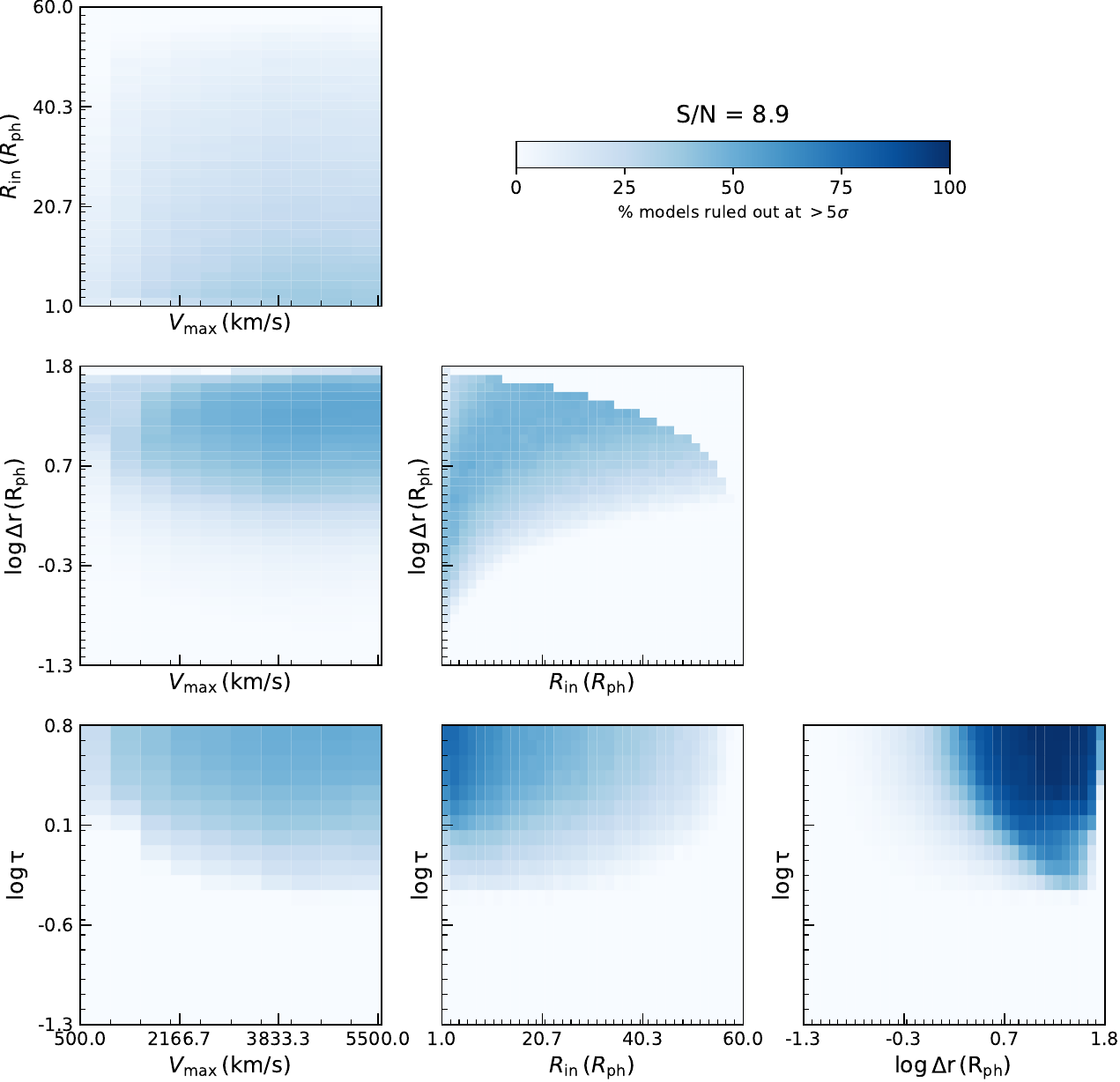}}
     \subcaptionbox{SN\,2022acch}{\includegraphics[width=0.45\textwidth]{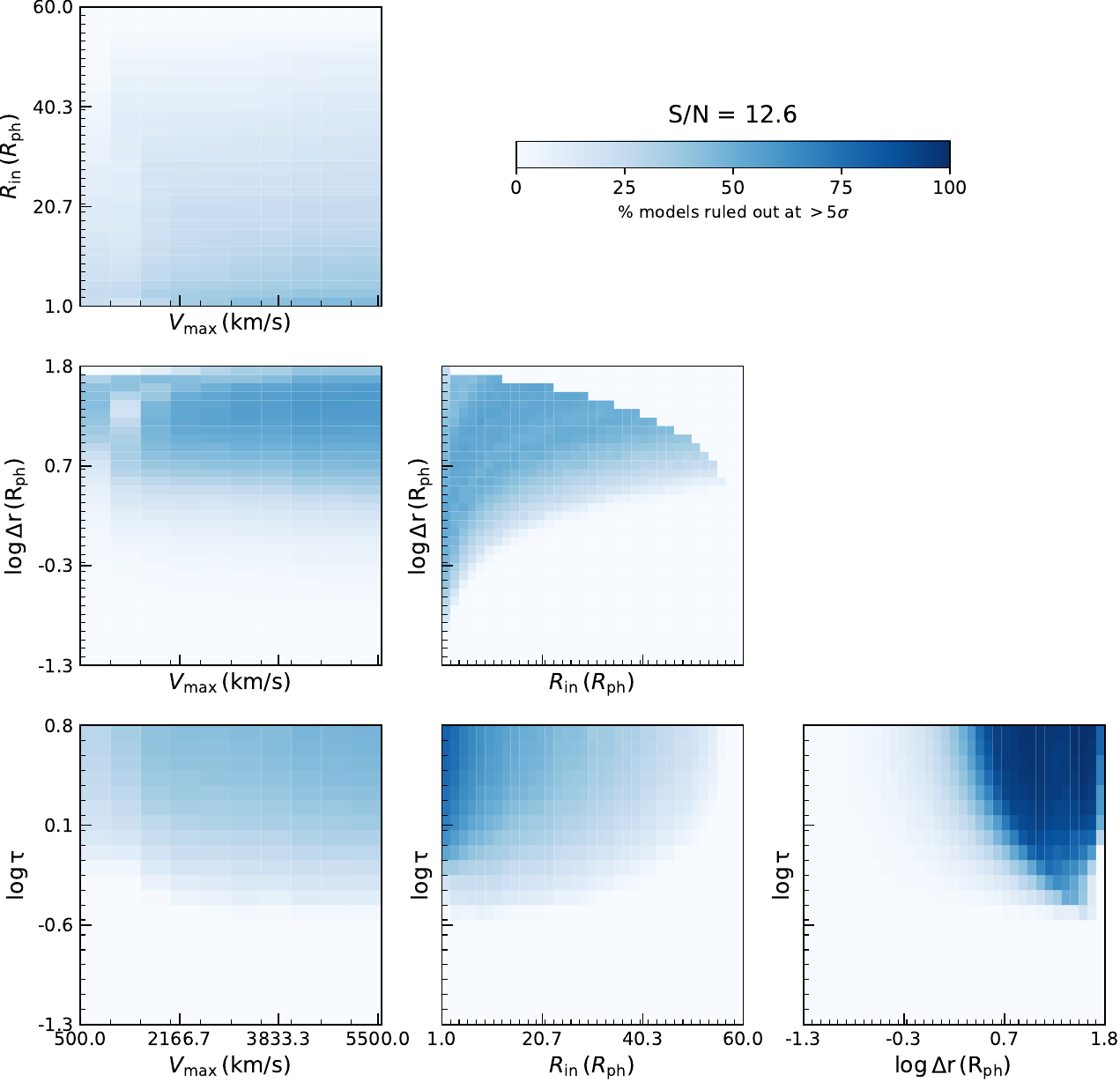}}
    \subcaptionbox{SN\,2020abjx}{\includegraphics[width=0.45\textwidth]{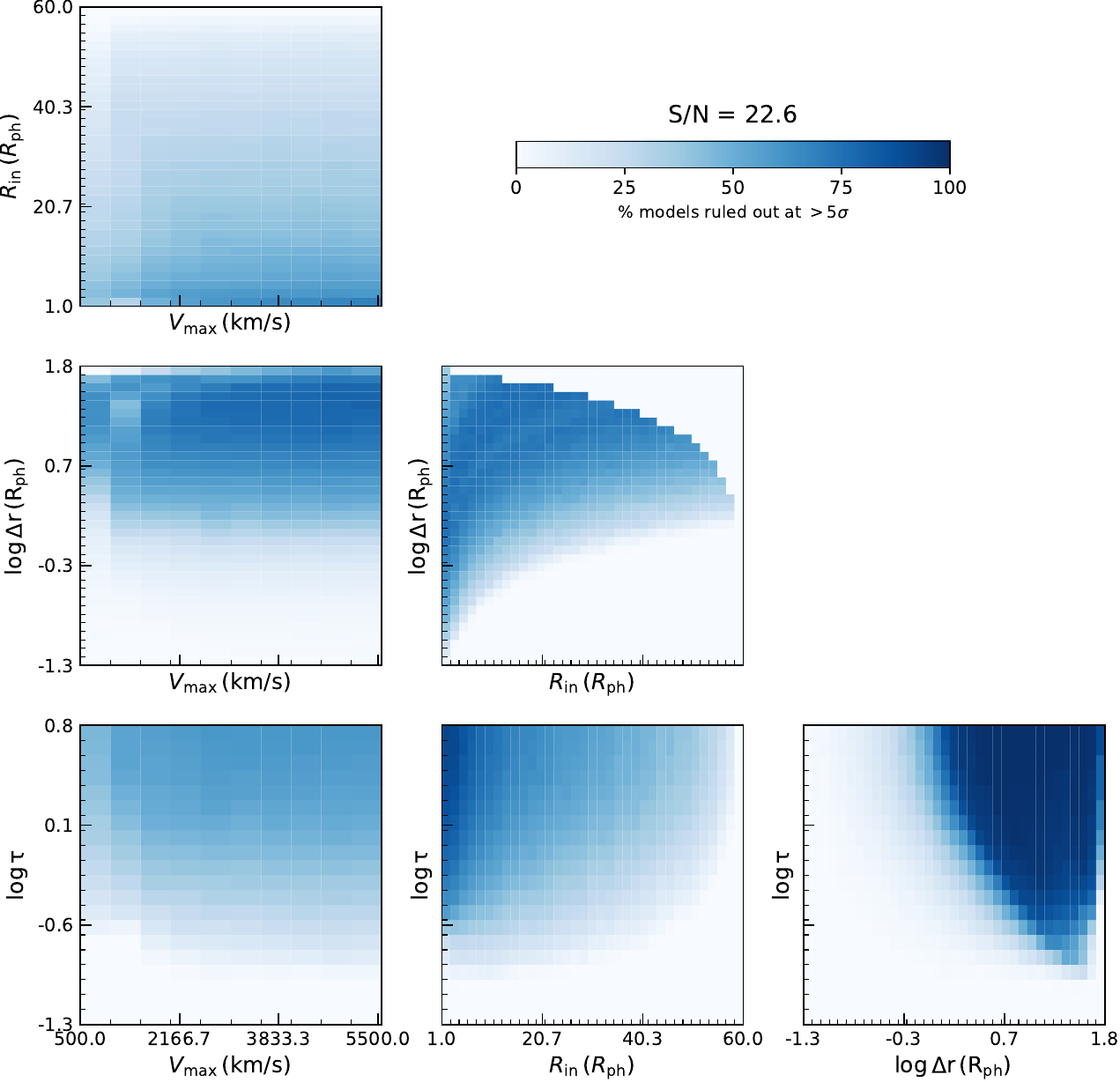}}
    \caption{Corner plots for four of the 16 events in our sample without CSM detections, illustrating the percentage of CSM models ruled out as a function of an increasing S/N. The colorbar indicates the fraction of excluded models. The parameters correspond to the maximum velocity $v_{\rm max}$, inner radius $R_{\rm in}$, shell thickness $\Delta r$ and optical depth $\tau$.}
    \label{fig:non_dete_1}
\end{figure*}

To assess the statistical significance of the \ion{Mg}{II} CSM features predicted by the Monte Carlo code and to evaluate which regions of parameter space are ruled out from the observed spectrum at a given confidence level, we employed a Bayesian framework. In this framework the log-likelihood $\log L$ is given by:

\begin{equation}
\log L = -\frac{1}{2}\sum_{i=1}^{N}
\left[ \frac{(y_i - \mu_i)^2}{\sigma_i^{2}} +\ln~\!\bigl(2\pi \sigma_i^{2}\bigr)
\right]
\label{eq:logL}
\end{equation}
where $y$ is the observed data, $\mu$ is a given model and $\sigma$ is the uncertainty in the observed data, under the assumption that the observational errors are independent and Gaussian distributed with known variances.

First, we defined the underlying continuum by fitting the observed spectrum with a spline function. We applied a $3\sigma$ clipping to remove residual artifacts and cosmic rays. To ensure that potential weak CSM features would not bias the continuum fit, we then interpolated across regions corresponding to host galaxy \ion{Mg}{II} absorption and the expected CSM feature locations (which is different for each model). To assess the effect of observational noise on the continuum fit, we applied a Monte Carlo approach. We generated multiple realizations of the observed spectrum by perturbing it according to the measurement uncertainties and for each realization, the continuum was refit. For each continuum, we computed the $\log L$ using Eq.~\ref{eq:logL}, where $\mu$ represents the continuum model. This process generates a distribution of $\log L_{\rm continuum}$ values that represent how well a continuum-only model can explain the observations under the observed noise.

Second, the synthetic spectra, generated by the Monte Carlo code, were degraded to match the spectral resolution of the X-shooter instrument, using the arm in which the \ion{Mg}{II} feature is observed. As these synthetic spectra assumed a flat continuum (see Sect.~\ref{sec:montecarlo_code}), they were then multiplied by the underlying continuum defined in the first step, producing realistic model spectra that include possible CSM absorption features. Using Eq.~\ref{eq:logL}, we computed the $\log L_{\rm feature}$, where $\mu$ now corresponds to the model with CSM shell (feature model), quantifying how well the feature model reproduces the observed data.

Third, to provide a measure of how strongly the data support the continuum model relative to the absorption model, we calculated:

\begin{equation*}
    \Delta \log L =\frac{ \langle \log L_{\rm continumm} \rangle - \log L_{\rm feature}} {\sigma \log L_{\rm continuum}}
\end{equation*}
which directly indicates the confidence level of the detection. We adopt a $5\sigma$ threshold to determine whether the model with a CSM shell is preferred over the continuum-only model (i.e., no CSM shell). In cases where the significance exceeds $+5\sigma$, the continuum-only model is strongly favored by the observed data and the corresponding CSM absorption models are ruled out. Conversely, when the significance falls below $-5\sigma$, the model with a CSM is significantly preferred over the continuum-only model, indicating the detection of a CSM shell in the observed spectrum. For significance values between $-5\sigma$ and $+5\sigma$, we cannot distinguish between the models and cannot rule out or confirm the presence of a CSM shell.

For objects in which the model with a CSM shell is never favored by the data, and thus no CSM absorption is detected in the observed spectra, we show the excluded parameter space for four objects in Fig.~\ref{fig:non_dete_1}. Similar diagnostic plots for the other 12 objects without detections are shown in Appendix~\ref{app:model_results}. The results are shown as a function of increasing S/N to illustrate how the excluded parameter space evolves with improving data quality. While no combination of the four parameters can be entirely ruled out at any S/N level, certain combinations are clearly disfavored. The most notable result is that models characterized by both large CSM shell thickness and high optical depth are strongly disfavored. As shown in Figs.~\ref{fig:parameter_grid_tau} and \ref{fig:parameter_grid_rout}, such configurations would produce strong and deep \ion{Mg}{II} absorption features that would almost certainly have been detected in all objects. Additionally, shells with high optical depth and small inner radius (i.e., located close to the photosphere) tend to be excluded more readily. This is expected, as shown in Fig.~\ref{fig:parameter_grid_rin}; nearby shells give rise to broad P-Cygni-like profiles with prominent emission components, which, when combined with high optical depth, result in highly conspicuous spectral features.

With increasing S/N, we are able to place progressively stricter constraints on the properties of potential CSM shells, as fewer configurations can remain hidden within the noise of the spectrum. As shown in Fig.~\ref{fig:non_dete_1}, for events with low S/N, the ruled-out parameter space is relatively limited, and much of the model space remains viable due to the limited ability of low-quality spectra to constrain subtle \ion{Mg}{II} absorption features.  At high S/N, the vast majority of the parameter space is excluded, permitting only configurations involving shells that are both optically and geometrically thin, i.e., those producing weak features that could plausibly remain undetected even in high-quality data.

\subsubsection{Physical constraints for undetected CSM shells}

\begin{figure}[!ht]
   \centering
  \includegraphics[width=0.5\textwidth]{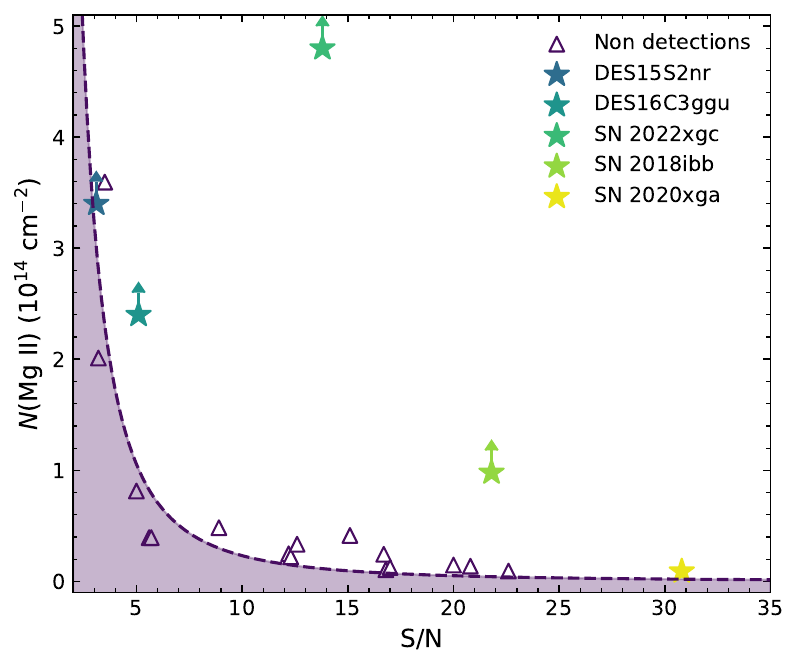}
   \caption{Column density of the CSM shells as a function of the spectrum’s S/N. Triangular markers denote the lower limits for objects with no detected shell, representing the column density above which a shell would have been detectable given the noise level. The dashed line shows the best-fit relation, and the shaded region marks the parameter space where a CSM shell would remain undetectable at the given S/N. Star symbols indicate the measured column densities (or lower limits) for objects with detected CSM shells.}
      \label{fig:col_den}
\end{figure}

To constrain the physical properties of the CSM shells that are ruled out, we estimate lower limits on their column densities using Eq.~\ref{eq:density}, above which the shells would have been detectable. In Fig.~\ref{fig:col_den}, we plot these column density limits as a function of the spectral S/N. As expected, the column density threshold shows an exponentially declining trend with increasing S/N, indicating that higher-quality data enable the detection of weaker CSM shells. This plot further illustrates that shells with properties similar to those of SN\,2018ibb, SN\,2022xgc, DES16C3ggu, and DES15S2nr would be detectable for S/N > 5 per resolution element, whereas weaker shells, such as that observed in SN\,2020xga, would require S/N > 25 per resolution element for detection.

Determining corresponding limits on the mass of these shells, which would provide stronger constraints on the nature of the ejections, remains challenging in the case of non-detections. Since no \ion{Mg}{II} absorption systems are observed, there are no specific shell parameters to anchor a mass estimate; the same column density could correspond to a range of shell masses depending on the assumed shell geometry.

We also investigated whether two distinct populations exist in the inferred ejection times, depending on whether the CSM shell is detectable or not. In principle, such a distinction could provide clues as to whether shells expelled immediately prior to explosion are swept up by the SN ejecta, while those located at somewhat larger radii remain observable. However, no clear separation is evident among the non-detection cases. Thus, the detectability of a shell likely depends on its density; shells ejected only a few days before explosion could be detectable if sufficiently dense, but would remain unseen if too optically thin. Moreover, even a dense but geometrically thin ($\Delta r \ll 1$) shell would be quickly overtaken by the SN ejecta, erasing its spectral signature within days. Similarly, shells expelled decades to hundreds of years prior may be observable if dense enough, but diffuse, low-density shells would produce no detectable absorption features at the time of observation.

\subsubsection{CSM shell detections}

\begin{figure}
    \centering
    {\includegraphics[width=0.5\textwidth]{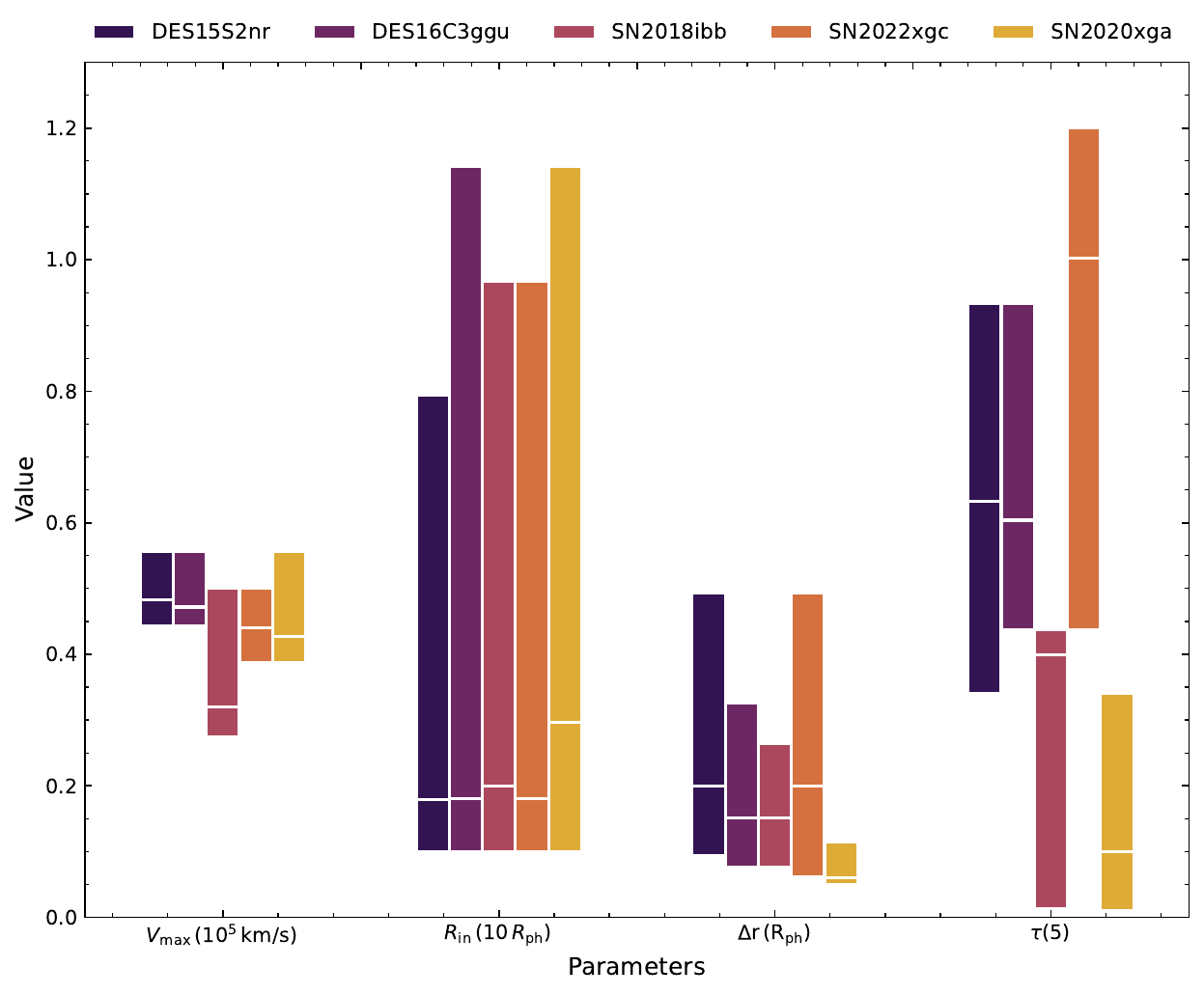}}
    \caption{Boxplots for DES15S2nr, DES16C3ggu, SN\,2022xgc, SN\,2018ibb, and SN\,2020xga. The colored boxes indicate the parameter space where models with CSM are statistically favored over continuum-only models at a confidence level exceeding $>5\sigma$. The white horizontal lines within each box mark the best-fit values. The parameters explored are the maximum velocity ($v_{\rm max}$), the inner radius ($R_{\rm in}$), the shell thickness ($\Delta r$), and the optical depth ($\tau$).}
    \label{fig:dete}
\end{figure}

We identified five objects (DES15S2nr, DES16C3ggu, SN\,2018ibb, SN\,2020xga and SN\,2022xgc) for which some of the models with a CSM shell are statistically preferred over the continuum-only models by the data with $>5\sigma$ significance. The favored regions of the parameter space are shown in Fig.~\ref{fig:dete}. In three of these events, SN\,2018ibb, SN\,2020xga, and SN\,2022xgc, previously discussed by \cite{Gkini2025}, broad \ion{Mg}{II} absorption features have been directly observed at velocities of $3200 - 4400~\mathrm{km~s^{-1}}$. As shown in Fig.~\ref{fig:dete}, the models favored by our analysis fall within these velocity ranges, consistent with the derived shell properties from \cite{Gkini2025}.

In this work, we also identify DES15S2nr and DES16C3ggu in which there are models with CSM shells that are favored by the data. In both cases, the preferred models lie in the velocity range of $\sim 4000-5000~\rm km~s^{-1}$ suggesting the presence of a CSM shell within this velocity range. Indeed, inspection of the spectra of DES15S2nr and DES16C3ggu reveals a second \ion{Mg}{II} absorption system at velocities close to $5000~\rm km~s^{-1}$. This confirms that our method has successfully identified two more objects exhibiting fast-moving CSM at $>5\sigma$ confidence level. We further note that the permitted velocity range is tightly constrained relative to the other parameters, as it dictates the specific spectral region where the CSM absorption is located. 

We note that the boxplots presented in Fig.~\ref{fig:dete} illustrate only the presence of models with CSM shells that are favored over continuum-only models in the observations. They do not imply that these models provide good absolute fits to the observed CSM absorption lines. This is because our method is designed as a model comparison tool to detect the existence of CSM features in the observed spectrum and not to find the best-fit model. Given the coarse, gridded nature of the parameter space and the constraints on computational time, extracting the exact shell properties of an observed CSM line is not feasible with the current approach. Incorporating a full Bayesian inference framework could enable a more rigorous parameter estimation, but implementing such a methodology is a future work focusing on the code development. Accurate measurements of the CSM properties therefore require a dedicated line-profile modeling method, which we discuss further in Sect.~\ref{sec:modelling_detections} for DES15S2nr and DES16C3ggu.

\subsection{CSM shells around DES15S2nr and DES16C3ggu} \label{sec:modelling_detections}

\begin{figure*}[!ht]
     \centering
          \begin{subfigure}[b]{0.49\textwidth}
         \centering
         \includegraphics[width=\textwidth]{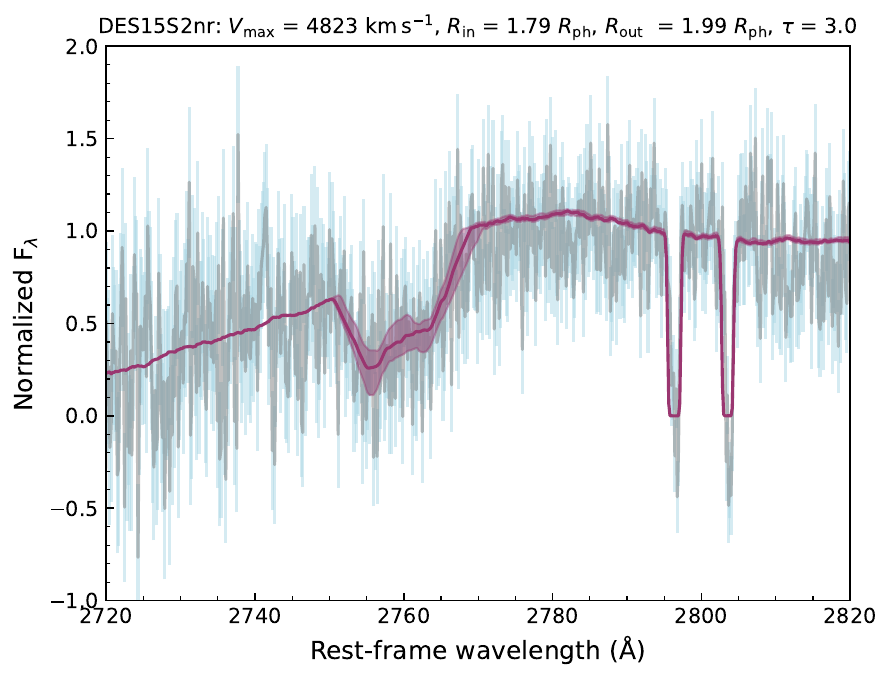}
     \end{subfigure}
     \begin{subfigure}[b]{0.49\textwidth}
         \centering
         \includegraphics[width=\textwidth]{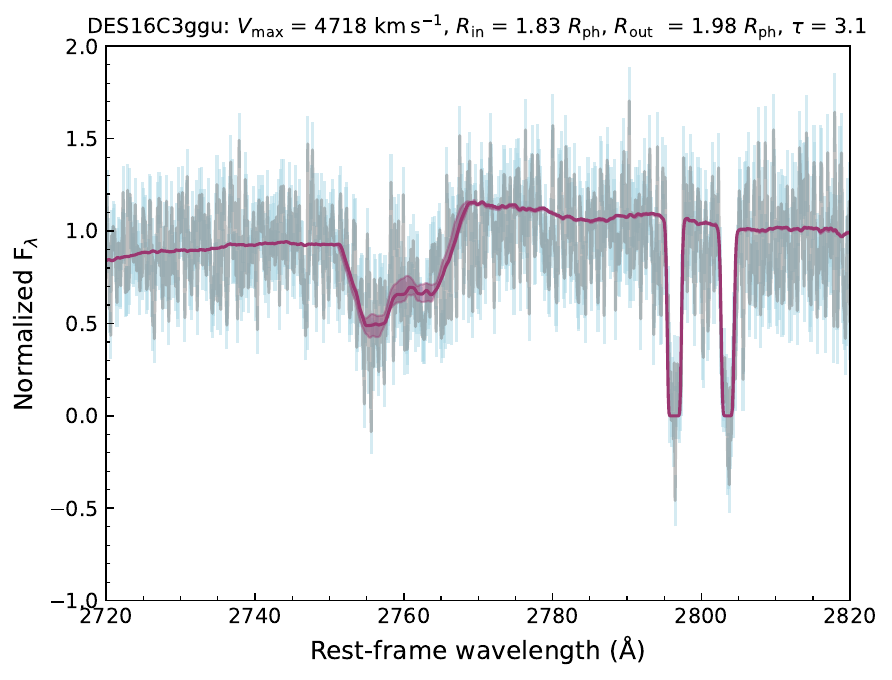}
     \end{subfigure}

    \caption{Modeling of the \ion{Mg}{II} doublets originating from the CSM shell (broad features) and the host galaxy’s interstellar medium (narrow features) for DES15S2nr (left) and DES16C3ggu (right). The observed spectra in the 2800~\AA\ region are shown in gray, while the best-fit model and associated $1\sigma$ uncertainty range are overplotted in purple. The titles of each panel indicate the CSM shell parameters corresponding to the best-fit model.}
    \label{fig:MgII_modelling}
\end{figure*}

We modeled the CSM shells of the two new objects, DES15S2nr and DES16C3ggu, for which our method indicated the presence of a CSM absorption feature. To refine the best-fit parameters, we first limited the parameter space to the regions in which  CSM models are favored over the continuum models. We then employed a Monte Carlo sampling approach to generate 10\,000 models centered around this region of parameter space. After determining the maximum-likelihood model from this first sampling, we further narrowed the parameter space and repeated the process, generating an additional 10\,000 models within the refined region. To estimate the $1\sigma$ uncertainty region, we selected all models whose likelihood values fell within $1\sigma$ of the best fit. The observed spectra, along with the best-fit models and $1\sigma$ associated uncertainties, are presented in Fig.~\ref{fig:MgII_modelling}.

For DES15S2nr, we find good agreement between the model and the observed spectrum for a shell with inner radius $R_{\rm in} = 1.79^{+0.32}_{-0.27}~R_{\rm ph}$ and outer radius $R_{\rm out} = 1.99^{+0.38}_{-0.30}~R_{\rm ph}$, where $R_{\rm ph}$ is the photospheric radius. The best-fit maximum velocity of the CSM shell is $v_{\rm max} = 4823^{+27}_{-120}~\rm km~s^{-1}$. These values are within the permitted region in Fig.~\ref{fig:dete}. From the observed spectrum, corrected for Milky Way extinction and calibrated against photometric data, we derived a blackbody radius of $1.88 \pm 0.08 \times 10^{15}~\rm cm$. Using this, we estimate that the CSM shell is located at $3.37^{+0.62}_{-0.53} \times 10^{15}~\rm cm$ and extends to $3.74^{+0.73}_{-0.59} \times 10^{15}~\rm cm$. 
Based on the derived shell location and velocity, we estimate that the shell was ejected $90.0^{+17.6}_{-14.2}$ days prior to the epoch of the observed spectrum. As no well-constrained non-detections exist prior to the first detection of the SN, we adopt the first detection date (MJD $57252.3$; \citealt{DAndrea2015}) as a proxy for the explosion time. Assuming a constant expansion velocity (as inferred for SN\,2018ibb), we estimate that the CSM shell was expelled approximately $2.0^{+0.6}_{-0.4}$ months before the core collapse.

For DES16C3ggu, the best-fit model corresponds to a CSM shell with inner and outer radii of $R_{\rm in} = 1.83^{+0.09}_{-0.07}~R_{\rm phot}$ and $R_{\rm out} = 1.98^{+0.10}_{-0.07}~R_{\rm phot}$, respectively, and a maximum expansion velocity of $v_{\rm max} = 4718^{+42}_{-16}\rm~km~s^{-1}$. These values are consistent with the regions where models with CSM are favored by the data. These values translate into a physical shell that extends from $4.81^{+0.24}_{-0.19} \times 10^{15}~\rm cm$ to $5.21^{+0.27}_{-0.19} \times 10^{15}~\rm cm$. Assuming that the shell has not experienced significant deceleration, we estimate that it was ejected $3.4^{+0.2}_{-0.1}$ months prior to the SN's first detection (MJD 57763.1; \citealt{Angus2019}), based on the inferred expansion velocity.

We emphasize that the uncertainties in the estimated ejection time for DES15S2nr and DES16C3ggu reflect only the statistical uncertainties from the spectral modeling. Systematic uncertainties, such as those arising from the poorly constrained explosion date or the blackbody fit to the spectrum which is notably noisy, are not accounted for. This is particularly severe for DES16C3ggu, which lies at a higher redshift and for which early-time photometric coverage is limited. Consequently, for both DES15S2nr and DES16C3ggu, we treat the estimated time between CSM ejection and the SN explosion as an upper limit. Nonetheless, the derived timescales are consistent with the scenario in which the CSM is expelled within the final year before the core collapse in SLSNe-I with detected CSM shells detected.

The relative depth of the blue and red \ion{Mg}{II} doublet in both DES15S2nr and DES16C3ggu indicate an optical depth of $\tau \approx 3$. Using Eq.~\ref{eq:density} we estimate the column density of the CSM \ion{Mg}{II} of DES15S2nr and DES16C3ggu. Since the lines are saturated in both cases, we only find a lower limit of $N(\ion{Mg}{II})  > 3.2 \times 10^{14}~\rm cm^{-2}$ and $N(\ion{Mg}{II})  > 2.4 \times 10^{14}~\rm cm^{-2}$ for DES15S2nr and DES16C3ggu, respectively.

For DES16C3ggu, only a single X-shooter spectrum was obtained, owing to the limited brightness and larger distance of the event. In the case of DES15S2nr, additional X-shooter spectra were acquired at later epochs; however, the UV region of these data 
is too noisy to reliably detect any potential signatures from the CSM shell.

\section{Light curve properties} \label{sec:phot_distribution}

\subsection{Detections versus non-detections}

\setlength{\tabcolsep}{9pt}

\begin{figure}[!ht]
   \centering
  \includegraphics[width=0.5\textwidth]{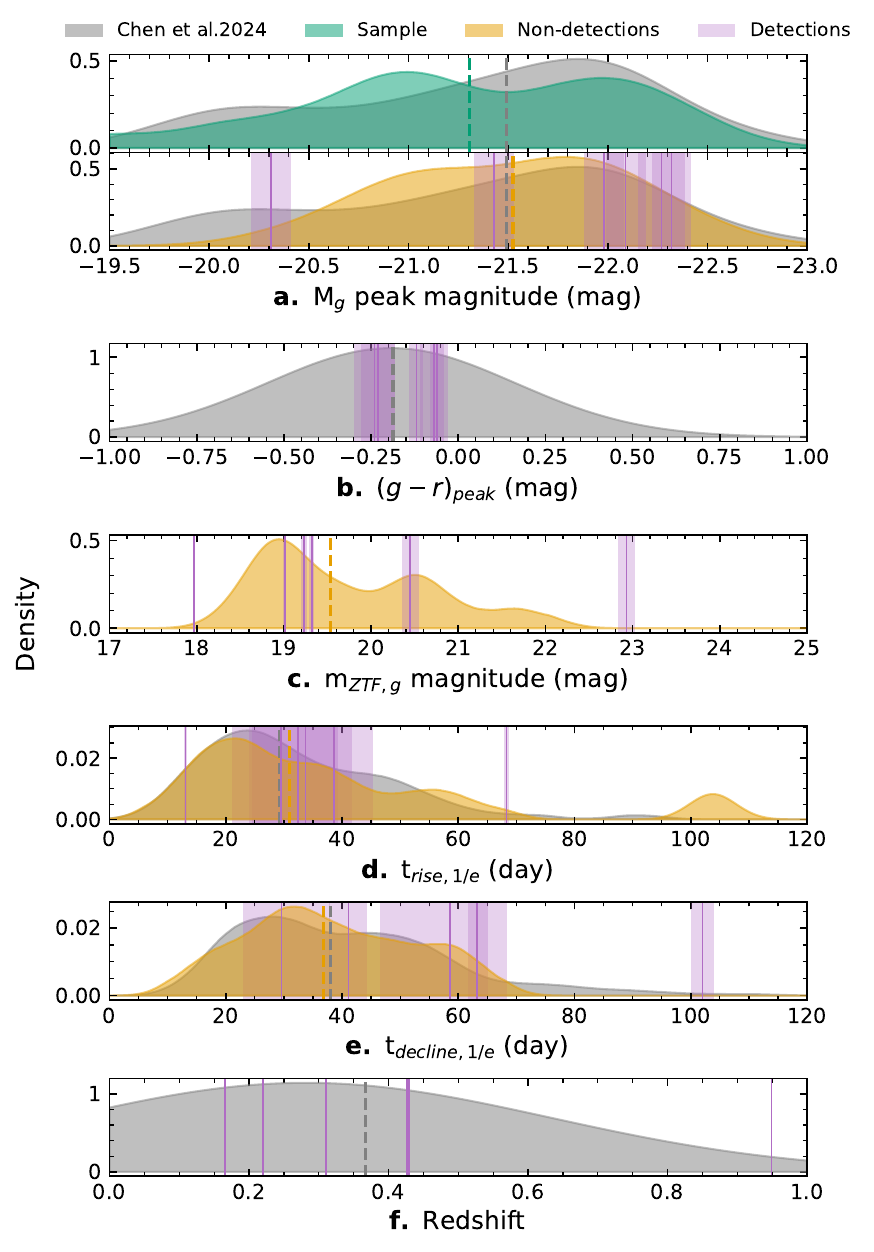}
   \caption{Comparison of the photometric properties of the final X-shooter sample with the ZTF SLSN-I sample from \citet{Chen2023a}. Panel a (upper): KDE distributions of $M_{g}$ peak magnitudes for 78 ZTF SLSNe-I and the triggered X-shooter sample. Panel a (bottom): KDE distributions of $M_{g}$ peak magnitudes for the ZTF sample, with the X-shooter detections and non-detections shown separately. Panel b: KDE distribution of $(g - r)_{\rm peak}$ for the ZTF sample, overlaid with individual measurements and $1\sigma$ uncertainties (purple lines and shaded regions) for the detection events. Panel c: KDE distributions of observed $m_{g}$ magnitudes for the non-detection subsample and individual measurements of the detection events. Panels d and e: KDE distributions of the rise and decline timescales for the ZTF and non-detection subsamples, along with individual measurements for the detection events. Panel f: KDE distribution of redshifts for the ZTF sample, with overlaid redshift values of the detection events. Vertical dashed colored lines indicate the median of each distribution.}
      \label{fig:photo_comp}
\end{figure}

To estimate the light curve properties in the rest-frame $g$ band for the SLSNe-I in our sample, we selected the appropriate observed photometric bands that correspond to the rest-frame $g$ band at various redshifts. Specifically, we used $g$-band light curves for objects with $z < 0.17$, $r$ band for $z > 0.17$, $i$ band for $z > 0.56$, and $z$ band for $z > 0.86$. The peak of each rest-frame $g$-band light curve was estimated using the method described by \citet{Angus2019}, which involves light-curve interpolation and Gaussian Process regression, implemented with the \texttt{GEORGE} Python package \citep{Ambikasaran2015} using a Matern 3/2 kernel. To convert the peak apparent magnitudes to peak absolute magnitudes, we applied the relation $M = m - \mu - A_{\rm MW} - K_{\rm corr}$, where $m$ is the apparent magnitude, $\mu$ is the distance modulus, $A_{\rm MW}$ is the Milky Way extinction, and $K_{\rm corr}$ is the K-correction. For the K-correction, we used the formula from \cite{Hogg2002} along with spectra near peak. In objects where peak spectra were unavailable, we applied the term $-2.5 \log(1+z)$, which has been found to agree within $0.1$~mag of the full K-correction \citep{Chen2023a}. The resulting rest-frame $g$-band absolute magnitudes are reported in Table~\ref{tab:lc_properties}.

In Fig.~\ref{fig:photo_comp}, we compare the light curve properties of our final X-shooter sample to those of the homogeneous ZTF SLSN-I sample presented by \citet{Chen2023a}, which analyzed the photometric characteristics of 78 H-poor SLSNe-I. Each panel displays the kernel density estimates (KDEs) for the ZTF sample, derived from a Monte Carlo simulation that incorporates the asymmetric uncertainties in the measured parameters. First, to assess whether our initial triggered sample selection introduces a bias toward intrinsically brighter SNe, we compare the peak magnitudes of our initially triggered sample to those of the ZTF sample in Fig.~\ref{fig:photo_comp}a (top panel). We find that the median peak magnitude of our initial sample, $-21.3 \pm 0.1$ (the mean absolute deviation value is reported), is consistent with the ZTF median of $-21.5 \pm 0.2$, indicating that our selection does not preferentially favor brighter SNe. Next, we divide the final X-shooter sample into events with and without \ion{Mg}{II} CSM detections, shown in Fig.~\ref{fig:photo_comp}a (bottom). The median peak magnitude of the non-detection subsample, $-21.5 \pm 0.3$, is consistent with that of the ZTF sample. For the detection subsample, which contains only six objects, we do not plot a KDE, as it would not reliably reflect the underlying distribution. Instead, we present the individual measurements along with their associated uncertainties. On average, the detection subsample appears somewhat more luminous than both the non-detections and the ZTF sample. However, two events with a CSM from the DES sample exhibit peak magnitudes that are fainter than the ZTF median, suggesting that the presence of detectable CSM absorption features is not strictly correlated with the intrinsic brightness of the SNe.

A comparative analysis of the temperature evolution between the detection and non-detection subsamples reveals no significant differences between the two groups. However, since this analysis was based solely on three optical photometric bands and the blackbody peak lies in the UV, we also use the rest-frame $g-r$ color at peak brightness as an additional diagnostic. To estimate this, we used the rest-frame peak magnitudes inferred from the interpolated rest-frame $g$- and $r$-band light curves. The resulting values are reported in Table~\ref{tab:lc_properties}. As shown in Fig.~\ref{fig:photo_comp}b, the $g - r$ colors at peak are consistent with the median of $-0.2 \pm 0.1$~mag the ZTF sample, supporting our earlier conclusions regarding the temperature evolution. Due to the relatively higher redshifts of the events in our X-shooter sample and the lack of observations in redder photometric bands and/or spectra at peak, this analysis could only be conducted for the detection events where data are available.

We investigated whether a CSM shell is more likely to be detected in spectra of SLSNe with a brighter apparent magnitude. For objects with ZTF photometry available at the time of spectroscopic observations, we used the observed $g$-band magnitudes. For those lacking ZTF coverage, we derived the synthetic $g$-band magnitude by absolutely calibrating the spectra, and convolving with the ZTF $g$-band transmission curve. The resulting observed magnitudes are listed in Table~\ref{tab:lc_properties}. As shown in Fig.~\ref{fig:photo_comp}c, the observed magnitudes of the detection events span the full range of the observed magnitude distribution, indicating no significant correlation between apparent brightness and the presence of CSM absorption features. Additionally, we investigated whether the detectability of CSM features correlates with the observed phase (relative to the time of first detection), under the hypothesis that such features may be more prominent at early phases before the SN ejecta overtakes the CSM shell. Our analysis revealed no clear trend between phase and detectability. Notably, in the case of SN\,2018ibb, the \ion{Mg}{II} absorption remains visible in a spectrum obtained approximately $127$ days after first detection, suggesting that such features can persist well beyond the early phase.

We used the interpolated rest-frame $g$-band light curves to define the rise and decline timescales based on fractional flux thresholds (e.g., $t_{\rm rise,1/e}$ is defined as the time interval between $f_{\rm peak}/e$ and $f_{\rm peak}$), following the methodology of \citet{Nicholl2015b}. All timescales are reported in rest-frame days. As shown in Fig.~\ref{fig:photo_comp}d and Fig.~\ref{fig:photo_comp}e, the median rise $31.1 \pm 12.2$~days and decline timescale $36.8 \pm 13.3$~days of the non-detection subsample are in agreement with that of the ZTF sample of $29.3 \pm 7.8$ and $38.0 \pm 12.1$, respectively. We note that five objects had insufficient light curve data (see Table~\ref{tab:lc_properties}) and were not included in this comparison. In contrast, the majority of the detection events exhibit, on average, longer rise times and significantly longer decline timescales compared to both the ZTF and the non-detection subsamples. These longer diffusion timescales may potentially reflect explosions of more massive progenitor stars and/or lower expansion velocities. However, a larger sample is required to draw statistically robust conclusions. Finally, in Fig.~\ref{fig:photo_comp}f, we compare the redshift distribution of the detection subsample to that of the ZTF SLSN-I sample. We find that the detections span the full redshift range of the ZTF sample, indicating that CSM interaction signatures are not confined to higher-redshift events. 

\begin{figure}[!ht]
   \centering
  \includegraphics[width=0.5\textwidth]{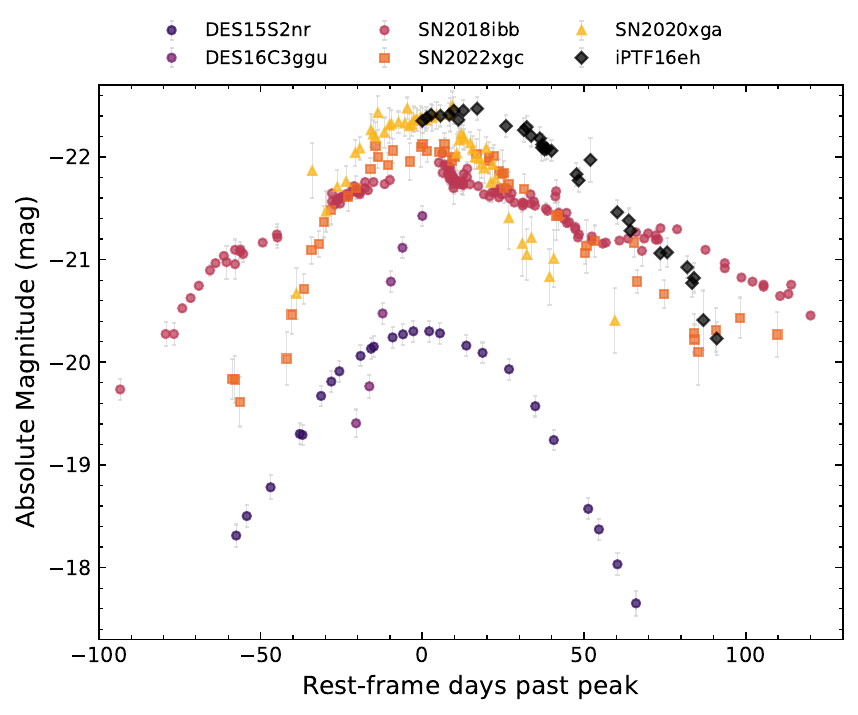}
   \caption{Rest-frame $g$-band absolute magnitude  light curves for the objects with detected CSM shells. The magnitudes are K-corrected and corrected for MW extinction. The x-axis is in rest-frame days with respect to the $g$-band peak, with the exception of iPTF16eh, where the $u$ band was utilized for estimating the peak owing to the lack of data in the rising part of the $g$-band light curve.}
      \label{fig:phot_csm_comp}
\end{figure}

To better illustrate the diversity in the photometric properties of the six objects exhibiting CSM interaction signatures, we present their rest-frame $g$-band light curves in Fig.~\ref{fig:phot_csm_comp}. All absolute magnitudes are K-corrected and corrected for Milky Way extinction. Among the sample, SN\,2018ibb stands out due to its markedly slow-evolving light curve, which features prominent bumps and undulations, in contrast to the smoother evolution observed in the other five SLSNe-I with \ion{Mg}{II} absorption systems. No signs of post-peak bumps or wiggles are evident in the light curves of SN\,2020xga, DES15S2nr, DES16C3ggu, and iPTF16eh, while SN\,2022xgc shows a possible flattening in the $gcr$ bands beginning approximately 80 days after peak brightness (see \citealt{Gkini2025}). Additionally, SN\,2022xgc exhibits a clear pre-peak bump in the rest-frame $g$ band shortly after explosion, and SN\,2020xga shows a possible early-time bump at around $-30$ days. We note that DES15S2nr does exhibit a pre-peak bump immediately after first detection in the rest-frame $u$-band light curve, as discussed by \citet{Angus2019} and interpreted within the shock breakout model for an extended CSM shell proposed by \citet{Piro2015}. However, the current data are insufficient to draw definitive conclusions regarding the origin of such pre-peak bumps in SLSNe-I. Moreover, such bumps are not observed in all events with CSM detections, as \citet{Angus2019} rule out the presence of a pre-peak bump in DES16C3ggu down to limiting magnitudes of $M \sim -16$~mag.

\subsection{SN properties versus CSM properties}

\begin{figure}[!ht]
   \centering
  \includegraphics[width=0.47\textwidth]{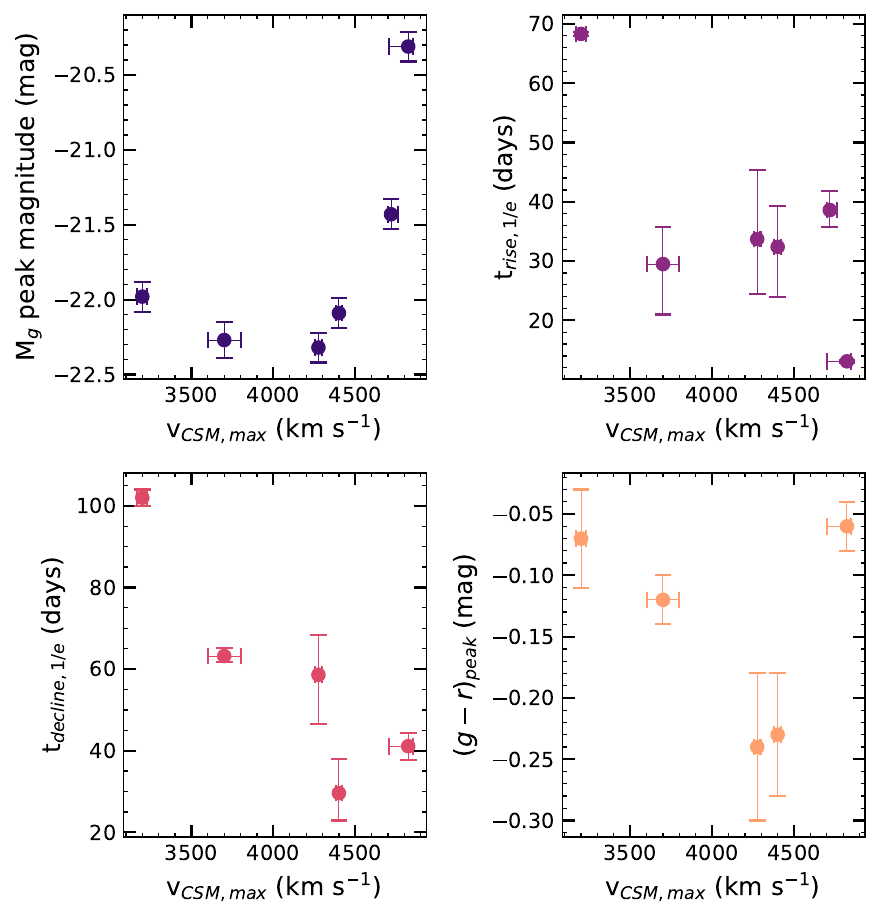}
   \caption{Photometric supernova properties as a function of the maximum CSM shell velocity. Panels show absolute $g$-band peak magnitude (top left), decline timescale (top right), rise timescale (bottom left), and $(g-r)$ color at peak (bottom right).}
      \label{fig:CSMvsSN}
\end{figure}

To investigate potential correlations between the SN and CSM properties, we compared the photometric characteristics of the six detections, including peak magnitude, rise and decline timescales and color at peak, with the inferred CSM parameters such as maximum velocity, radius, shell thickness and column densities. Owing to the absence of extensive multi-band photometric coverage and the wide range of redshifts of the detected events, estimates of peak bolometric luminosities and radiated energies do not yield meaningful constraints. Furthermore, the lack of high-quality spectroscopic data around peak light, combined with the limited spectral coverage for several of the objects, prevents us from examining potential connections between spectroscopic SN properties (e.g., ejecta velocities) and the inferred CSM shell properties.

In Fig.~\ref{fig:CSMvsSN}, we show the maximum velocity of the CSM shells, as inferred from our modeling of the detection sample, plotted against key photometric properties of the SNe. We do not find any correlation between the CSM velocity and the peak magnitude, rise time or color at peak. Similarly, no correlation is observed between the SN properties and the CSM radius, thickness, or column density. However, we identify a possible negative correlation between the CSM maximum velocity and the decline timescale, with a Spearman coefficient of $\rho = -0.90$ and a null-hypothesis probability of $p = 0.04$. If genuine, this trend could suggest that SLSNe-I with faster-declining light curves tend to host faster-moving CSM shells, potentially linking the SN explosion to the preceding mass-ejection episode. We caution, however, that this is a simplistic picture. The decline timescale may also be influenced by other factors, such as the powering mechanism of the SLSNe. Given the limited number of detections (six in total), these results cannot be considered statistically robust. A larger sample will be required to establish whether any meaningful connection exists between the ejection of CSM shells and the subsequent SN explosion.

\section{Discussion} \label{sec:discussion}

\subsection{The mass of the detected CSM shells} \label{sec:mass}

To estimate a lower limit on the mass of the CSM shells which have been detected both in this work and in \cite{Gkini2025}, from the optical depth of \ion{Mg}{II}, we adopt the Sobolev approximation (Eq.~\ref{eq:tau}) for a homologous expanding medium and solve for the mass of the shell $M_{\rm shell}$:

\begin{equation*}
    \frac{M_{\rm shell}}{\rm M_{\odot}} \approx \frac{\tau}{6.5 \times 10^{12}} \left(\frac{v_{\rm max}}{1000~\rm km~s^{-1}}\right)^{3}  \left(\frac{t}{\rm month}\right)^{2}\frac{\mu}{X(\rm Mg {\rm II})}
\end{equation*}
where $t$ is the time since ejection, $X(\ion{Mg}{II})$ is the \ion{Mg}{II} abundance by number, and $\mu$ is the mean molecular weight of the dominant element in the shell.

Assuming that these shells were ejected very close to the explosion, they are likely among the last shells expelled and therefore dominated by heavier elements (such as O). To set a lower limit on the shell mass, we used the \ion{Mg}{} abundances predicted in the PPI ejection models for H- and He-poor material of \cite{Yoshida2016} for the most massive progenitor stars, corresponding to a number of $X(\rm Mg II) \approx 3.5 \times 10^{-2}$ (see their Fig.~11). Under the assumption that all Mg is present as \ion{Mg}{II}, we derive shell masses in the range of $10^{-8} - 10^{-6}~\rm M_{\odot}$. Although these values should be considered strict lower limits, they are still very small and inconsistent with the shell masses predicted by PPI models. 

The discrepancy between these estimated mass limits and theoretical predictions may arise from several factors. First, the detected shells might not be O-dominated but instead H- or He-dominated. To explore this scenario, we adopted the solar abundance of Mg ($X(\rm Mg II) \approx 4 \times 10^{-5}$ by number) as a proxy and derive shell masses of of $10^{-6} - 10^{-5}~\rm M_{\odot}$. While these values are two orders of magnitude higher than those estimated from an O-dominated shell, they remain low. In addition, as noted by \citet{Schulze2024} and \citet{Gkini2025}, the shells are unlikely to be H-dominated, since H must have been removed prior to these eruptions. This interpretation is further supported by SN\,2018ibb, where the absence of H and He lines throughout the spectral evolution, combined with the presence of prominent O lines, suggests that the CSM shell is indeed O-dominated, with any residual H or He likely confined to much larger radii.

Alternatively, some of the other assumptions may be invalid, such as spherical symmetry of the shell, the adopted Mg abundance, or the ionization state of Mg (e.g., a significant fraction in \ion{Mg}{III} or higher). The latter is particularly important (if not dominant) since the ionization can be strongly influenced by extreme-UV and X-ray emission. This is a natural consequence of any circumstellar interaction model, including PPI eruptions \citep[e.g.,][their fig. 13]{Tolstov2017}. In this scenario, the inferred shell mass could increase by several orders of magnitude. Thus, in the absence of a complete physical model, estimates of the shell mass remain highly uncertain and depends on the underlying mass-ejection mechanism.

\subsection{The origin of the CSM shells} \label{sec:origin_shells}

In Sect.~\ref{sec:model_nondete}, we reported the detection of CSM shells in DES15S2nr and DES16C3ggu at $5\sigma$ confidence levels.  These findings increase the number of SLSNe-I with detected CSM shells from four (as of 2024) to six. However, a key open question remains: what physical mechanism drives the ejection of CSM shells to the inferred radii and velocities, given that the material appears to have been expelled only a few months to less than a year prior to core collapse.

Motivated by the inferred ejection velocities and timescales, we examine potential ejection mechanisms within the framework of eruptive mass loss. One plausible scenario for the origin of the detected CSM shells could be giant eruptions similar to those seen in luminous blue variable (LBV) stars. The best-studied case is the ``Great Eruption'' of $\eta$ Carinae in 1843, during which $12$–$20~\rm M_{\odot}$ (or possibly more; \citealt{Smith2003,Morris2017}) of material was expelled over the course of a decade, with ejecta velocities up to $20000~\rm km~s^{-1}$ \citep{Davidson2001,Currie2002,Smith2002,Smith2004,Smith2008,Smith2018}. Observations of LBV nebulae, such as that around $\eta$ Carinae, show that the geometry of the ejecta can be highly anisotropic. Although the physical mechanisms driving LBV eruptions remain uncertain, several possibilities have been proposed, including envelope instabilities, binary interactions, and wave-driven mass loss \citep[e.g.,][]{Davidson1987,Owocki2004,Smith2003,Smith2006,Smith2008,Woosley2017,Akashi2020,Cheng2024}. For events such as iPTF16eh, SN\,2020xga, and SN\,2022xgc, \citet{Lunnan2018} and \cite{Gkini2025} suggested that an LBV-like eruption could not be excluded as the origin of the observed CSM shell. Nonetheless, \citet{Lunnan2018} highlighted that the detached, roughly spherical shell observed in iPTF16eh is inconsistent with the typically asymmetric CSM morphology associated with LBV eruptions, which tend to produce material spanning a broad range of velocities (see Fig. 1 in \citealt{Smith2008}). In addition, the CSM masses inferred in Sect.~\ref{sec:mass} are significantly smaller than that observed in $\eta$ Carinae. Despite these inconsistencies, the inferred shell velocities of $\sim 5000~\rm km~s^{-1}$, combined with the high uncertainties in the CSM mass estimates and the poorly constrained shell geometry, prevent us from definitively ruling out eruptive mass-loss events analogous to those observed in LBV eruptions.

The LBV-like eruptions have also been proposed as potential outcomes of the PPI mechanism \citep{Woosley2017}. He stars with masses between $30$–$65~\rm M_{\odot}$ can experience recurrent pair-instabilities that can lead to the ejection of massive shells of material.  In these models, the key parameter that governs the time interval between shell ejections and the final collapse, is the He core mass $M_{\rm He}$ \citep{Woosley2017,Leung2019,Marchant2019,Huynh2025}. As discussed by \citet{Gkini2025}, we consider the PPI scenario primarily as a mechanism capable of producing the observed CSM shells, rather than as the power source of the SLSNe themselves. We compare our results with the PPI models because these currently provide the only theoretical predictions available in the literature that can be directly confronted with the inferred properties of the detected shells, particularly the shell velocities, which fall within the relatively narrow range predicted by PPI models. Indeed, SN\,2018ibb, which is considered the best pair-instability candidate, cannot be be placed within the PPI framework. Instead, \citet{Schulze2024} discussed this event in the context of an LBV-like eruption, analogous to $\eta$ Carinae.

A comparison with PPI models from \cite{Woosley2017}, \cite{Marchant2019}, \cite{Renzo2020} and \cite{Huynh2025} can set a lower limit on the stellar mass, assuming that the models must have at least one pulse on a timescale equivalent to the ejection time of the CSM shell we observed. For DES15S2nr and DES16C3ggu, given that the CSM shells were expelled $\sim 2$ and $\sim 3.5$ months, respectively, before the first detection, the mass of the progenitor is estimated to be $>45~\rm M_{\odot}$. For SN\,2020xga and SN\,2022xgc, the corresponding estimates are $>47 ~\rm M_{\odot}$ and $>46~\rm M_{\odot}$. These limits could be higher if the observed shell ejections correspond to the second pulse \citep{Renzo2020,Huynh2025}. The differences between these estimates arise from variations in the treatment of shocks and convection \citep{Leung2019} in the different models, as well as from differing assumptions regarding metallicity, binarity, rotation and other stellar parameters.

We emphasize that our comparison with PPI models is focused solely on the properties of the ejected CSM shells, and we do not attempt to directly link these models to the light-curve or spectral characteristics of the SNe with detected shells. Nevertheless, several SLSNe-I discussed in the literature in the context of PPI, as well as PPI models \citep{Tolstov2017, Woosley2017,Huynh2025} exhibit smooth light curves without prominent bumps or undulations, while showing timescales broadly consistent with those inferred for the objects with detected CSM shells.
Moreover, for several of the SLSNe-I with detected CSM shells analyzed in this work, the light curves are poorly constrained and only a single spectrum is available, limiting our ability to determine whether signatures of late-time shell-ejecta interaction might emerge and thereby provide tighter constraints on the shell ejection mechanism. Thus, an ejection mechanism analogous to PPI remains a plausible origin for the observed shells -- although it is not considered as a power source in this study -- and a qualitative comparison with PPI models provides useful constraints on the progenitor masses. Comparisons with future predictions from binary evolution models would be a valuable complement, as binary interactions represent an additional channel that may produce CSM shells shortly before explosion \citep[e.g.,][]{Laplace2021,Laplace2025a, Fang2025}. However, current binary models do not predict mass ejection at velocities comparable to those observed in our objects. Determining the chemical composition of the detected CSM shells from, for example, UV observations of different absorption or emission lines, could also provide critical clues about their ejection mechanism, but current observations are insufficient to constrain this.

We therefore conclude that, although some open issues remain between currently available eruptive mass-loss models (e.g., PPI or LBV-like eruptions) and the observed properties of SLSNe, the inferred shell characteristics require an eruptive mass-loss mechanism, and existing models can reproduce several key aspects of the shell observations. In particular, the velocities and timescales of the eruptions in the PPI models are consistent with what we infer from the observations in this work. Alternatively, explaining both the SLSN and CSM shell properties may require a different physical scenario that is currently unexplored or unknown. While the present work cannot definitively identify the mechanism responsible for these ejections, these results underscore the need for more detailed theoretical models, capable of simultaneously accounting for both pre-explosion eruptions and the subsequent SN explosions.

\subsection{Intrinsic CSM shell properties versus observational biases}
 
When we compare the properties of the CSM shells in the objects with detections, we find that the shells are located at $R_{\rm in} \sim 1.8 - 2.9~R_{\rm ph}$, with thicknesses of $\Delta r \sim 0.05 - 0.20~R_{\rm ph}$, and velocities spanning $3200 - 4800~\rm km~s^{-1}$. Although these shells are not identical, they display broadly similar characteristics, which may hint at a common physical origin. In particular, they all lie within a relatively narrow velocity range and are located at comparable radii. As illustrated in Fig.~\ref{fig:non_dete_1} the number of ruled-out models is similar across the full velocity grid, indicating that the detections are not biased toward specific velocities. This suggests that the presence of shells at these relatively high velocities is likely a genuine property rather than a selection effect. As discussed in Sect.~\ref{sec:origin_shells}, such high velocities may be indicative of an eruptive mechanism, capable of expelling such fast-moving shells. Indeed, \cite{Huynh2025} predict shell velocities reaching up to $4500~\mathrm{km~s^{-1}}$, consistent with the values inferred in the objects with CSM shells.

Figure~\ref{fig:non_dete_1} also shows that shells located closer to the photosphere are more frequently ruled out than those farther away, indicating that the apparent clustering of detected shells at small radii could reflect an observational bias. Likewise, all inferred ejection times are less than a year before the SN explosion. Since Sect.~\ref{sec:rule_out} shows no evidence for two distinct populations, we cannot determine whether this short ejection timescale is intrinsic to the eruption mechanism or primarily a selection effect. The main exception is iPTF16eh, whose distinct properties are discussed separately in Sect.~\ref{sec:light_echo}.

\subsection{When do we see a light echo?} \label{sec:light_echo}

Among the six objects in which CSM was detected, only iPTF16eh exhibited the emergence of \ion{Mg}{II} emission line between 100 and 300 days after maximum light. During this period, the line shifted from $-1600$ to $+2900~\rm km~s^{-1}$, and was attributed to a light echo from the CSM shell. This shell was located at $48.1~R_{\rm ph}$ and inferred to have been ejected approximately 30 years prior to the SN explosion. Its properties therefore lie outside the parameter range derived for the other five objects with CSM shells. We note that the shell characteristics of iPTF16eh were not constrained through the absorption-line modeling that was carried out in \cite{Gkini2025} and this work, as the absorption features were unresolved at the instrumental resolution, but instead from modeling the temporal evolution of the emergent \ion{Mg}{II} emission line.

The unique case of iPTF16eh raise the question of under what conditions a light echo becomes observable. Light echoes occur when radiation emitted during the explosion is scattered by surrounding material and redirected toward the observer, arriving with a time delay due to the longer light path. The condition for observing an echo is that the characteristic duration of the exciting radiation satisfies $t_{\rm rad} \ll R_{\rm shell} / c$. This requirement was clearly met in the best-studied SN\,1987A, where the shock breakout, lasting only minutes, served as the source of radiation for the CSM ring located at a distance of $\sim 200$ light days, resulting in a clearly observed echo. In this case the ring expansion was $\sim 10~\rm kms^{-1}$ and thus, the wavelength shift with time was marginal \citep{Lunqvist1996}. For iPTF16eh, the shell was located at $\sim 120$ light days, and the light curve declined by $\sim 2~\rm mag$ in less than 100 days. Under these conditions, a light echo was detectable, and the higher shell velocity allowed the wavelength shift to be tracked over $\sim 200$ days.

For the remaining five objects with CSM, the shells inferred from our modeling are located at distances of $1.5$–$5$ light days, which is significantly shorter than the duration of the light curves. As a result, no observable effects of a light echo would be expected in these cases. Nevertheless, the presence of additional CSM shells at larger radii ($\gtrsim 100$ light days) could, in principle, produce observable light echoes, if the emission lines were brighter than the background continuum. Such distant shells would give rise to \ion{Mg}{II} emission features, whose wavelength shifts with time would depend on the shell velocity. However, the lack of high-quality late-time observations, comparable to those obtained for iPTF16eh, prevents us from confirming the presence of such features in our detection sample. The only object with late-time spectral coverage is SN\,2018ibb. \cite{Schulze2024} analyzed the spectra of SN\,2018ibb taken between $+230$ and $+378$ days after peak brightness and reported a \ion{Mg}{II} emission. Yet, due to significant rebinning of the data, it remains uncertain whether this emission is associated with a light echo from a CSM shell. Consequently, we are unable to track the evolution of the \ion{Mg}{II} line profiles, which would otherwise provide valuable constraints on the geometry and kinematics of the CSM. To date, iPTF16eh remains the only SLSN-I for which a spectroscopic light echo has been firmly identified, offering a rare window into the structure and evolution of the surrounding material. This does not, however, exclude the possibility of additional CSM shells at distances comparable to iPTF16eh in the other five objects with CSM detections.

\subsection{Why do CSM shells remain undetected in most SLSNe?}

Among the 21  SLSNe-I modeled in this work, we find evidence for a \ion{Mg}{II} feature originating from a CSM shell in only five cases with more than $5\sigma$ significance. If we also include the detection in the case of iPTF16eh, which was not modeled here due to its unresolved \ion{Mg}{II} absorption doublet, the total number of SLSNe-I exhibiting this feature increases to six. This corresponds to an observed fraction of $\sim 30\%$ of the total sample. We emphasize, however, that this fraction does not represent an intrinsic occurrence rate, since no volume correction has been applied and not all SLSNe-I discovered to date have been studied with sufficient spectral coverage or resolution. Nevertheless, our findings highlight the open question of whether such features are intrinsic to all SLSNe-I but remain undetected due to observational limitations, or whether they reflect a genuine physical diversity in which only a subset of progenitors eject such material prior to explosion.

Our final X-shooter sample comprises spectra with S/N ranging from 3 to 32. For each object, this analysis allows us to place a lower limit on the column density of a CSM shell (see Fig.~\ref{fig:col_den}) above which an absorption feature would have been detectable given the noise level. Shells with properties comparable to those of SN\,2022xgc, SN\,2018ibb, DES15S2nr, and DES16C3ggu would have been detected in spectra with S/N $>5$, allowing us to exclude the presence of such shells in the remaining objects. However, the absorption lines detected in our sample are not identical across all objects, exhibiting variations in velocity, width, and strength. The most striking example is SN\,2020xga, in which our method yield a $>5\sigma$ detection, yet a shell with similar properties would only be detectable in spectra with S/N > 25 (see Fig.~\ref{fig:col_den}). As a result, non-detections do not necessarily imply the true absence of CSM; weaker lines arising from different CSM configurations may fall below the detection threshold.

Additional evidence for the presence of CSM around SLSNe-I comes from objects that show late-time H emission features in their spectra \citep{Yan2015,Yan2017,Pursiainen2022,Gkini2024}. These lines are interpreted as signatures of interaction with H-rich shells located at distances of $10^{15} - 10^{16},\rm cm$, likely expelled a few years prior to explosion \citep{Yan2017}. Among these objects, only SN\,2020zbf has spectral coverage around $2800$~\AA\ and was modeled in this work. However, no \ion{Mg}{II} absorption was detected, which may imply that the column density of the H-rich shell lies below our detection threshold. However, the existence of a few SLSNe-I with late-time H$\alpha$ emission clearly demonstrates that shells do exist around some SLSNe-I, even if they cannot always be revealed through the \ion{Mg}{II} absorption.

To explore whether the SLSNe-I in our work with detected CSM are unique, we performed a photometric analysis of the CSM-detected sample in comparison with the broader SLSN-I population. This comparison shows that objects with detected CSM shells do not stand out in terms of either peak brightness or light-curve timescales. Although most lie toward the more luminous end of the distribution, two objects are comparatively less luminous, indicating that the presence of CSM is not restricted to the brightest events. Furthermore, no clear correlation is observed between the vast majority of the SN properties and their inferred CSM characteristics. A possible trend between CSM velocity and light-curve decline time, which could provide insights into the progenitor, is hinted by the data; however, the limited sample size precludes drawing firm conclusions. These findings suggest that, the SN properties of our limited sample are insufficient to predict the properties of the CSM expected for a given SN. It therefore remains uncertain whether such CSM ejections occur in only a subset of progenitors or represent a more general phenomenon. However, the fact that the detected shells lie within a relatively narrow velocity range hints that a distinct physical mechanism may be responsible for driving these eruptions.

We conclude that the diagnostic developed in this work can identify detections of CSM features at $>5\sigma$ significance and place meaningful limits on CSM shells that are ruled out by the noise of the spectrum at the same confidence level. Weaker or intrinsically different CSM features (as in SN\,2020xga) may remain hidden below our observational limits, underscoring that detections require very high quality data. Although SLSN-I progenitors are expected to undergo mass ejections following the stripping of their H and He envelopes prior to explosion, the diversity of these ejections could account for the presence of absorption lines in the observed spectra. The timing of such ejections, the density and structure of the CSM shells may determine whether the features rise above the noise level of the spectra. The absence of a direct connection between the properties of the SNe and the inferred CSM characteristics prevents us from determining whether the observed fraction of SLSNe-I with detectable CSM shells reflects an intrinsic physical diversity among SLSN-I progenitors. Nevertheless, the detection of CSM-related features in six events demonstrates that at least a subset of SLSNe-I originate from massive progenitors that undergo substantial mass ejections, with velocities of $3000-5000~\rm km~s^{-1}$, occurring within months to years before core collapse. Identifying and characterizing such cases provides rare and valuable insights into the poorly understood final stages of massive stellar evolution.

\section{Conclusions} \label{sec:conclusion}
For this work, we presented a dedicated analysis of the near-UV spectra of 21 SLSNe-I to search for signatures indicative of CSM shells ejected shortly before core collapse, and we developed a diagnostic tool to quantify the detection significance of such CSM shells. Our key findings are summarized as follows:

\begin{itemize}
    \item Out of the 21 objects in our sample, five (SN\,2018ibb, SN\,2020xga, SN\,2022xgc, DES15S2nr, and DES16C3ggu) show robust detections of a CSM feature at the $>5\sigma$ level.  
    \item Modeling the broad \ion{Mg}{II} absorption indicates a CSM shell for DES15S2nr located at $\sim 3.4 \times 10^{15}\,\rm cm$ and for DES16C3ggu at $\sim 4.8 \times 10^{15}\,\rm cm$, expanding at maximum velocities of $\sim 4800~\rm km~s^{-1}$ and $\sim 4700~\rm km~s^{-1}$, respectively. These CSM shells were likely expelled only a few months prior to core collapse as a result of eruptive mass loss, although the exact mechanism cannot be securely determined.  
    \item Comparison with PPI models implies He-core masses $>45~\rm M_{\odot}$ for both DES15S2nr and DES16C3ggu.  
    \item The photometric properties of the detection sample are consistent with those of the broader SLSN-I population.
    \item We did not find correlations between the SN and CSM properties, except for a marginally significant correlation between the light curve decline time and the CSM shell velocity. However, a larger sample is needed to draw firm conclusions.
    \item The velocities of the detected shells are confined to a relatively narrow range of $\sim 3000-5000~\rm km~s^{-1}$, which may be indicative of a particular eruption mechanism.
    \item Shells with properties similar to those observed would have been detectable in spectra with a S/N $>5$, except in the case of geometrically and/or optically thin configurations. Therefore, the non-detections are unlikely to arise from selection effects and may instead suggest that only a subclass of SLSN-I progenitors experience such late-stage shell ejections shortly before explosion.
\end{itemize}

Looking ahead, facilities such as the Rubin Observatory’s Legacy Survey of Space and Time (LSST) will revolutionize the studies of SLSNe-I. LSST is expected to discover and monitor thousands of SLSNe-I, greatly expanding sample sizes and enabling robust, population-level analyses of progenitor diversity. Although high-quality UV spectroscopy will remain challenging for high-redshift objects, deeper photometric observations of nearby SLSNe-I could reveal precursor events and potentially link the CSM shells to the progenitor properties and explosion mechanisms. Furthermore, our study underscores the need for more accurate stellar and mass-loss models specifically tailored to the observations. Together, these developments will help disentangle observational limitations from intrinsic physical diversity, offering critical insights into the nature of SLSN-I progenitors.

\begin{acknowledgements}

AG thank GR for the valuable discussions regarding different aspects on this paper. 
Based on observations obtained with the Samuel Oschin telescope 48-inch and the 60-inch Telescope at the Palomar Observatory as part of the Zwicky Transient Facility project. ZTF is supported by the National Science Foundation under Grant No. AST-2034437 and a collaboration including Caltech, IPAC, the Weizmann Institute of Science, the Oskar Klein Center at Stockholm University, the University of Maryland, Deutsches Elektronen-Synchrotron and Humboldt University, the TANGO Consortium of Taiwan, the University of Wisconsin at Milwaukee, Trinity College Dublin, Lawrence Livermore National Laboratories, IN2P3, University of Warwick, Ruhr University Bochum, Cornell University, and Northwestern University. Operations are conducted by COO, IPAC, and UW.
The ZTF forced-photometry service was funded under the Heising-Simons Foundation grant \#12540303 (PI: Graham).
The Gordon and Betty Moore Foundation, through both the Data-Driven Investigator Program and a dedicated grant, provided critical funding for SkyPortal.

This work has made use of data from the Asteroid Terrestrial-impact Last Alert System (ATLAS) project. ATLAS is primarily funded to search for near earth asteroids through NASA grants NN12AR55G, 80NSSC18K0284, and 80NSSC18K1575; byproducts of the NEO search include images and catalogs from the survey area. The ATLAS science products have been made possible through the contributions of the University of Hawaii Institute for Astronomy, the Queen’s University Belfast, the Space Telescope Science Institute, and the South African Astronomical Observatory.

Funded by the European Union (ERC, project number 101042299, TransPIre). Views and opinions expressed are however those of the author(s) only and do not necessarily reflect those of the European Union or the European Research Council Executive Agency. Neither the European Union nor the granting authority can be held responsible for them.

AG and RL are supported by the European Research Council (ERC) under the European Union’s Horizon Europe research and innovation programme (grant agreement No. 10104229 - TransPIre).
U.B is funded by Horizon Europe ERC grant no. 101125877.
SS is partially supported by LBNL Subcontract 7707915.
MN is supported by the European Research Council (ERC) under the European Union’s Horizon 2020 research and innovation programme (grant agreement No.~948381).
NS and AS are supported by the Knut and Alice Wallenberg foundation through the “Gravity Meets Light” project.
T.-W.C. acknowledges financial support from the Yushan Fellow Program of the Ministry of Education, Taiwan (MOE-111-YSFMS-0008-001-P1), and from the National Science and Technology Council, Taiwan (NSTC grant 114-2112-M-008-021-MY3).
\end{acknowledgements}

\bibliographystyle{aa}
\bibliography{references}
%
%

\begin{appendix}

\FloatBarrier

\onecolumn

\section{Properties of the X-shooter sample} 

\begin{table}[!ht]
\centering
\caption{X-Shooter spectroscopic observations of the sample.}\label{tab:spectra_log}
\begin{tabular*}{.98\linewidth}[!ht]{@{\extracolsep{\fill}}lccllc}
\hline
\hline
Object & UT date & MJD & Phase\tablefootmark{\scriptsize a} & S/N\tablefootmark{\scriptsize b}  & Exposure  \\
& & (days) & (days) & & (s)  \\
\hline
SN\,2020xga & 20201107 & 59160.6 & $16.5$ & $30.8$ & 3600\\
SN\,2020abjx & 20201208 & 59191.5 & $74.5$ & $22.6$ &  3600  \\
SN\,2018ibb & 20190110 & 58493.1 & $126.5$ & $21.8$ &  1800  \\
SN\,2020zbf & 20201118 & 59170.8 & $28.5$ & $20.8$ & 2400   \\
OGLE15qz & 20151213 & 57369.1 & $18.5$ & $20.0$ &  2400   \\
iPTF13ajg & 20130417 & 56399.4 & $21.0$ & $17.0$ &  4800   \\
SN\,2020rmv & 20201015 & 59137.6 & $70.3$ & $16.8$  & 3600\\
SN\,2021hpx & 20210508 & 59342.6 & $52.6$ & $16.7$ &  2400  \\
SN\,2021ek & 20210116 & 59230.6 & $16.6$ & $15.1$  &  2400  \\
SN\,2022xgc& 20221207  & 59930.6 & $63.8$ & $13.8$ & 3600 \\
SN\,2022acch & 20221224 & 59937.8 & $27.3$ & $12.6$ & 2400   \\
SN\,2021gch & 20210416 & 59320.1 & $18.8$ & $12.3$ &  1500  \\
SN\,2020abjc & 20210313 & 59286.6 & $135.0$  & $12.2$ &   2400  \\
SN\,2022abdu & 20221215 & 59928.6 & $44.7$ & $8.9$ & 2400  \\
LSQ12dlf & 20120912  & 56182.5 & $50.7$ & $5.7$ &  1300   \\
SN\,2013dg & 20130626 & 56469.9 & $35.5$ & $5.6$ &  1300  \\
DES16C3ggu & 20170225 & 57809.0 & $23.6$ & $5.1$ &  4500   \\
DES16C3dmp & 20161221 & 57743.2 & $26.9$ & $5.0$ &  4500   \\
SN\,2021fao & 20210320 & 59293.8 & $31.2$ & $3.5$ &   2400  \\
iPTF15cyk & 20151013 & 57308.5 & $16.6$ & $3.2$ &  4800   \\
DES15S2nr & 20150919 & 57284.3 & $26.3$ & $3.1$ &  4500   \\ 

\hline
\end{tabular*}
\tablefoot{\tablefoottext{a}{Rest-frame days relative to the first detection.}\tablefoottext{b}{In the rest-frame region of 2675 -- 2875~\AA.}}
\end{table}

\begin{table*}[!ht]
\centering
\caption{Light curve properties of the X-shooter sample.} \label{tab:lc_properties}
\begin{tabular}{lcccccc}

\hline
\hline
Object & $M_{\rm g,peak}$  & $m_{\rm g}\tablefootmark{\scriptsize a}$ & Peak & $t_{\rm rise, 1/e}$ & $t_{\rm decline, 1/e}$ & $(g-r)_{\rm peak}$\\
 & (mag)  & (mag) & (J2000) & (days) & (days) & (mag)\\
\hline
SN\,2020xga & $-22.37 \pm 0.10$  & $19.32 \pm 0.03$ & $59172.5_{-11.7}^{+9.4}$ & $32.4_{-8.4}^{+6.9}$  &$29.6_{-6.8}^{+8.4}$ & $-0.23 \pm 0.05$\\
SN\,2020abjx & $-21.89 \pm 0.10$ & $19.55 \pm 0.03$ & $59218.2^{+6.7}_{-7.6}$ & $61.0^{+6.8}_{-7.8}$ & $44.3^{+17.2}_{-9.0}$ & --\\
SN\,2018ibb & $-21.80 \pm 0.02$ & $17.97 \pm 0.01$ & $58458.0^{+2.0}_{-2.0}$ & $68.3^{+0.4}_{-0.4}$ & $102.0^{+2.0}_{-2.0}$ & $-0.12\pm0.02$\\
SN\,2020zbf & $-20.96 \pm 0.01$  & $18.72 \pm 0.05$& $59164.8_{-1.0}^{+1.0}$ &$15.9^{+1.0}_{-1.1}$ & $41.9^{+2.0}_{-1.9}$ & $-0.16 \pm 0.12$\\
OGLE15qz & $-21.98 \pm 0.11$ & $ 20.41\pm 0.06 $ & $57373.2^{+3.5}_{-3.7}$ & $54.9^{+4.0}_{-4.0}$ &  $61.6^{+4.3}_{-4.3}$ & $-0.13 \pm 0.03$\\

iPTF13ajg & $-22.36 \pm 0.11$&  $20.70 \pm 0.09$ & $56406.8^{+3.0}_{-2.9}$ & $23.8^{+3.4}_{-3.2}$ & $36.0^{+3.1}_{-3.4}$ & -- \\
SN\,2020rmv & $-21.61 \pm 0.10$ & $18.94 \pm 0.02$ & $59117.4^{+7.6}_{-6.3}$ & $39.8^{+6.2}_{-5.6}$ & -- & -- \\
SN\,2021hpx & $-21.89 \pm 0.10$  & $18.50 \pm 0.04$ & $59335.3^{+3.0}_{-4.4}$ & $34.8^{+3.1}_{-4.4}$ & -- & --\\
SN\,2021ek & $-21.01 \pm 0.12$  & $18.80 \pm 0.03$ & $59237.4^{+5.2}_{-2.8}$ &  $20.9^{+5.4}_{-3.0}$ & $16.9^{+5.2}_{-2.9}$ & --\\
SN\,2022xgc & $-21.99 \pm 0.10$   & $19.23 \pm 0.03$ & $59901.9_{-12.0}^{+15.1}$ & $33.7_{-9.2}^{+11.6}$ & $58.6_{-12}^{+9.8}$ & $-0.24 \pm 0.05$\\
SN\,2022acch & $-22.26 \pm 0.12$ & $19.50 \pm 0.06$ & $59953.3^{+6.1}_{-7.5}$  & $38.1^{+6.2}_{-7.6}$ & -- & -- \\
SN\,2021gch & $-21.69 \pm 0.13$ &  $20.56\pm 0.07$ & $<59291.8$ & -- & -- & --\\
SN\,2020abjc &  $-21.09 \pm 0.03$ & $18.96 \pm 0.04$ & $59284.9^{+4.7}_{-5.1}$ & $103.8^{+1.4}_{-1.4}$ & $57.4^{+2.1}_{-2.1} $ & $-0.35 \pm 0.03$ \\
SN\,2022abdu & $-20.78 \pm 0.10$ & $18.96 \pm 0.09$ & $59914.8^{+5.0}_{-2.8}$ & $11.8^{+5.0}_{-2.9}$ & $22.5^{+5.3}_{-6.7}$ & --\\
LSQ12dlf & $-21.14 \pm 0.11$  & $20.44 \pm 0.05$ & $56138.2^{+2.4}_{-1.9}$ & -- & $30.6^{+2.1}_{-2.6}$ & --\\
SN\,2013dg & $-21.27 \pm 0.12$ & $19.84 \pm 0.06$ & $56449.1^{+4.4}_{-7.7}$ & -- & $28.5^{+7.8}_{-4.5}$ & $-0.18\pm0.08$\\
DES16C3ggu & $-21.43 \pm 0.11$ & $21.66 \pm 0.30$ & $>57803.0$ & $13.3_{-0.2}^{+0.2}$ & -- & --\\
DES16C3dmp & $-20.48 \pm 0.10$ & $21.83 \pm 0.20$ & $57743.0^{+2.8}_{-1.8}$  & $22.7^{+2.8}_{-1.8}$ &--& $-0.09\pm0.03$ \\ 
SN\,2021fao & $-21.89 \pm 0.10$  & $18.95 \pm 0.03$ & $59308.9^{+5.9}_{-7.4}$ & $29.3^{+6.0}_{-7.4}$  & $30.7^{+7.5}_{-6.0}$ & -- \\
iPTF15cyk & $-21.69 \pm 0.10$ & $21.35 \pm 0.30$ & $<57283.0$ & -- & $49.1^{+0.8}_{-1.0}$ & --\\
DES15S2nr & $-20.31 \pm 0.10$ & $20.45 \pm 0.10$ & $57322.5^{+3.5}_{-3.0}$ & $38.6^{+3.2}_{-2.9}$ &  $41.1^{+3.2}_{-3.3}$ & $-0.06 \pm 0.02$\\

\hline
\end{tabular}
\tablefoot{All measurements are reported in the AB system. \tablefoottext{a}{Observed magnitude at the time the spectra were taken.}}
\end{table*}

\onecolumn

\section{Spectroscopic modeling of the X-shooter sample} 

\label{app:model_results}

\begin{figure*}[h]
  \centering
    \subcaptionbox{iPTF15cyk}{\includegraphics[width=0.3\textwidth]{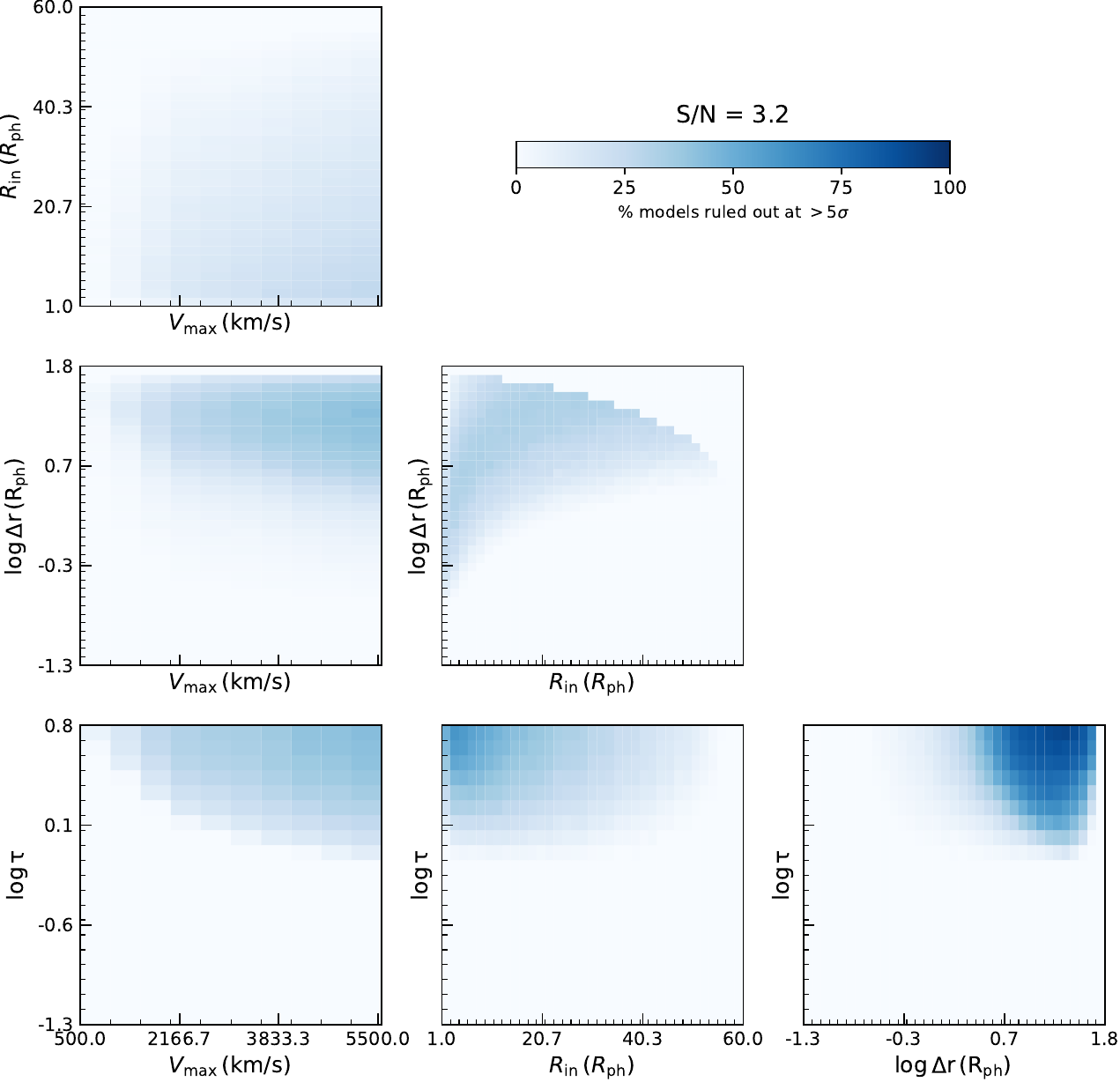}}
    \subcaptionbox{DES16C3dmp}{\includegraphics[width=0.3\textwidth]{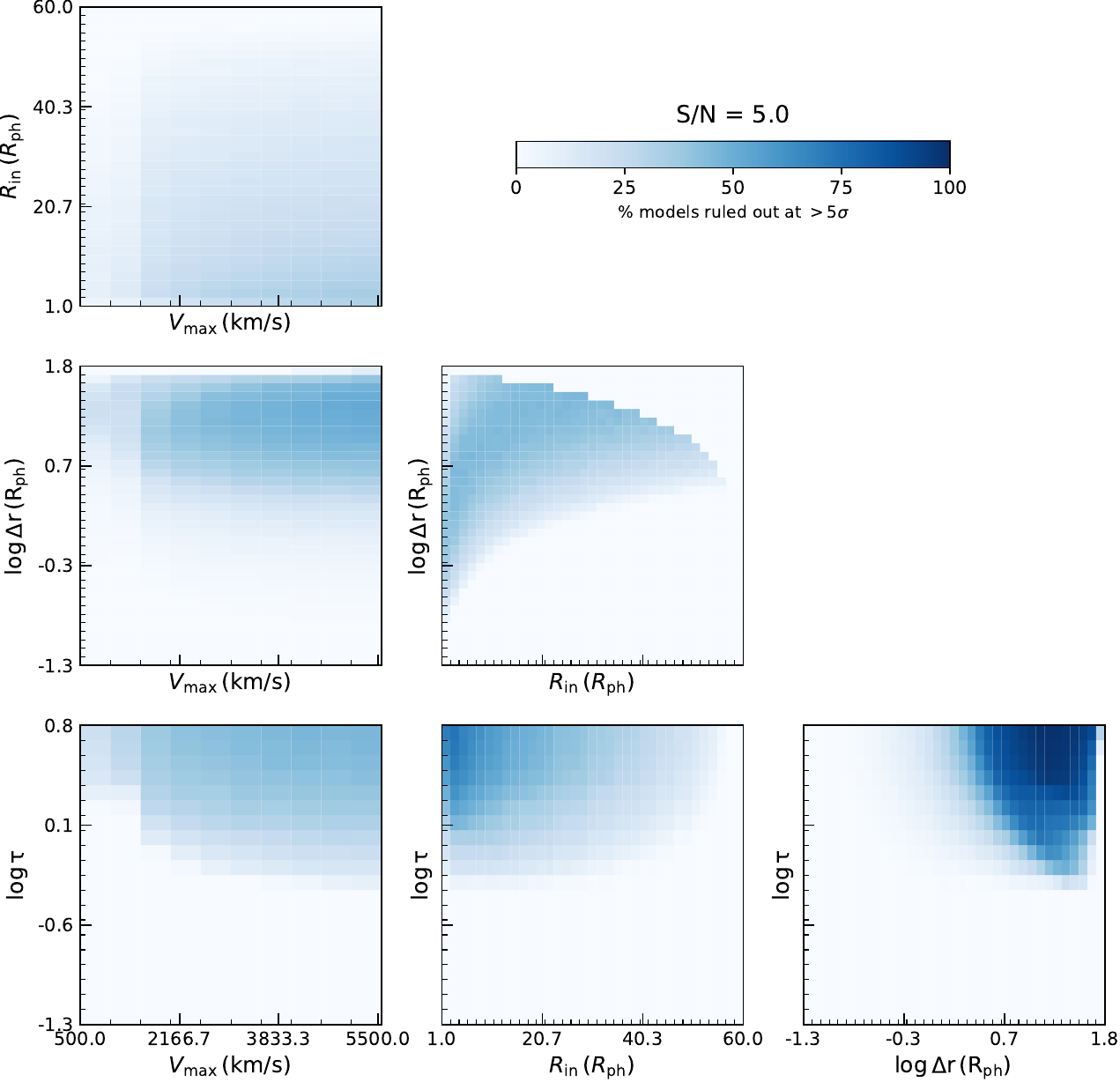}} 
     \subcaptionbox{SN\,2013dg}{\includegraphics[width=0.3\textwidth]{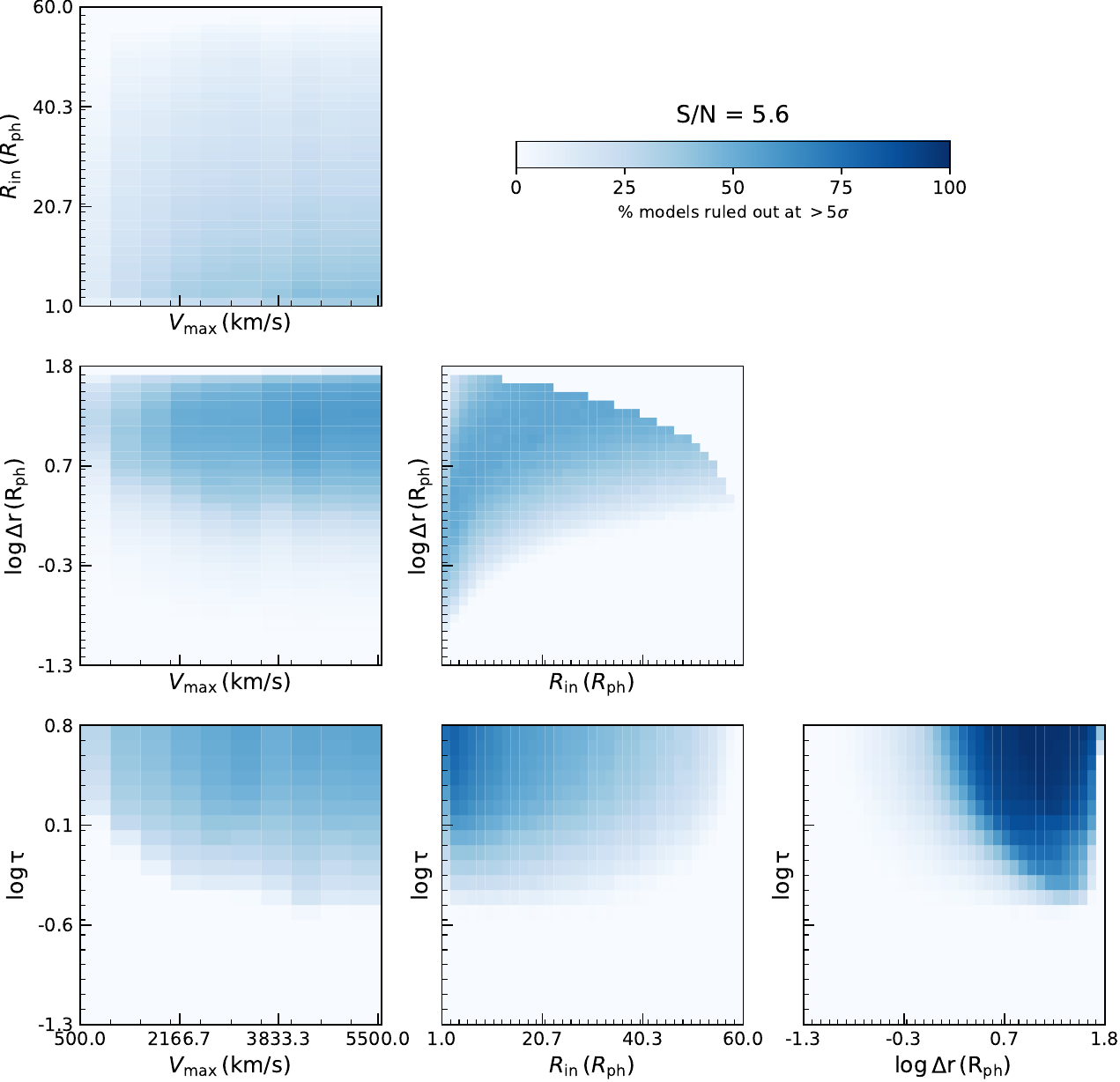}} \\
    \subcaptionbox{LSQ12dlf}{\includegraphics[width=0.3\textwidth]{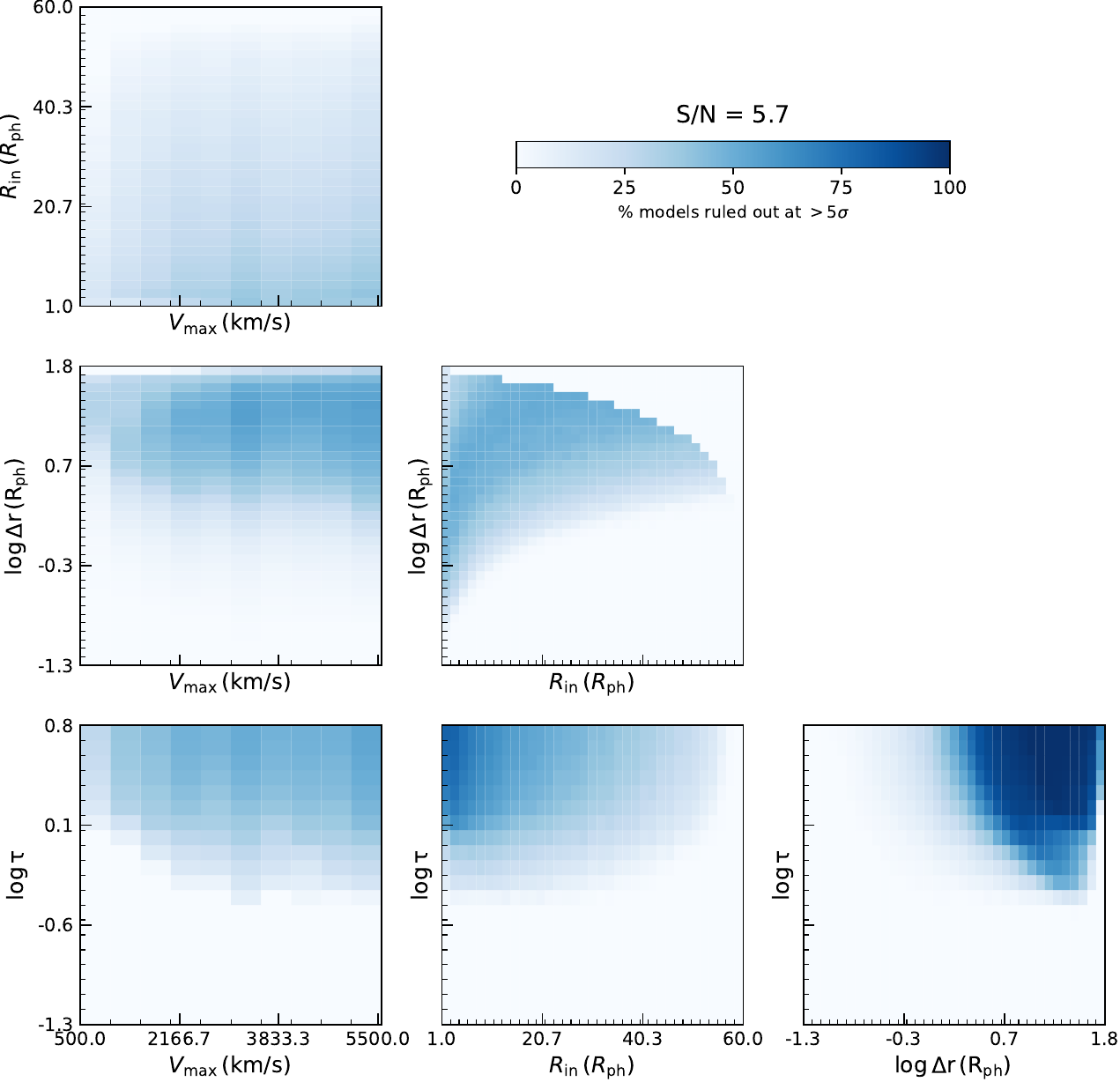}}
    \subcaptionbox{SN\,2020abjc}
    {\includegraphics[width=0.3\textwidth]{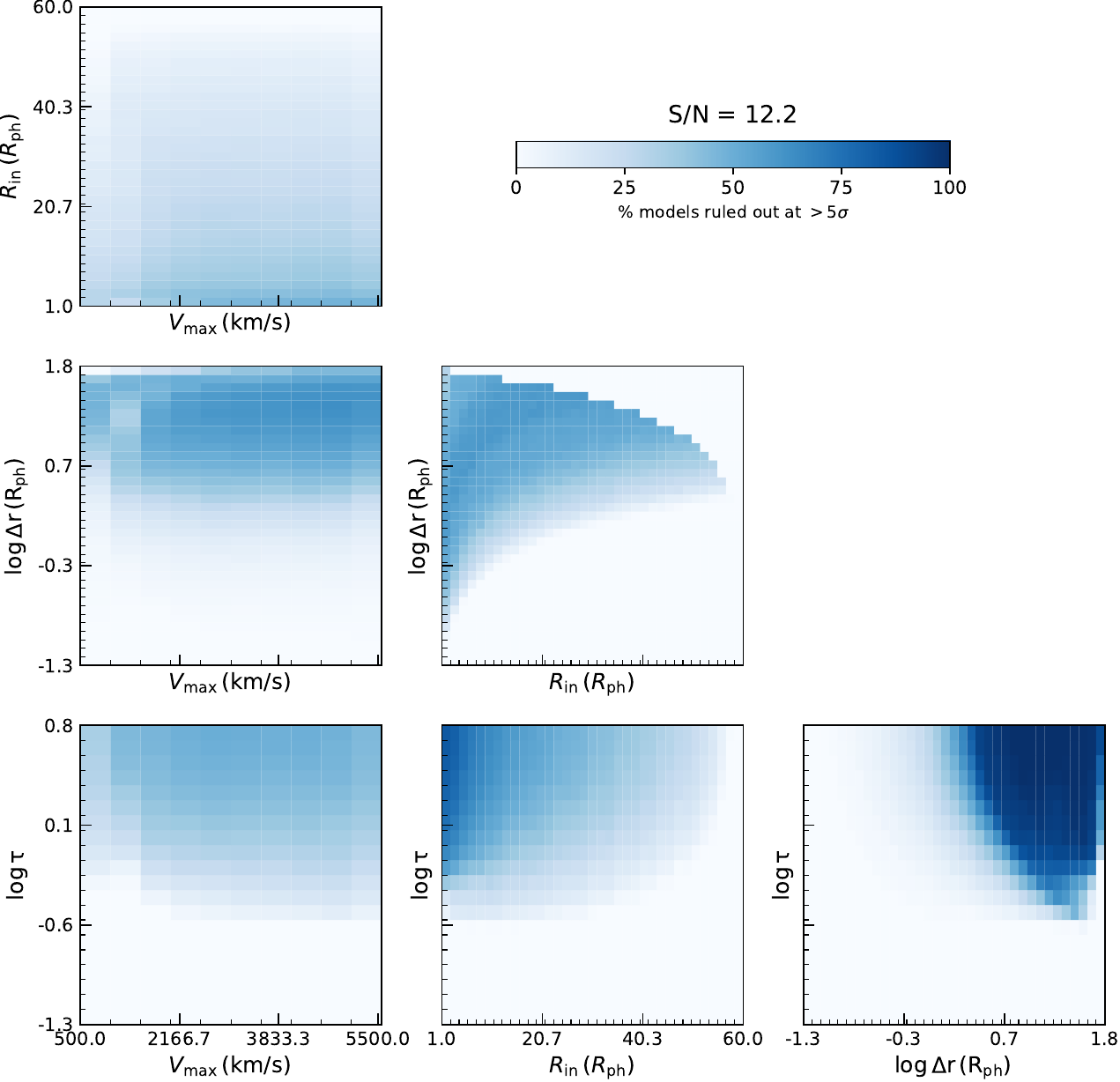}}
    \subcaptionbox{SN\,2021gch}{\includegraphics[width=0.3\textwidth]{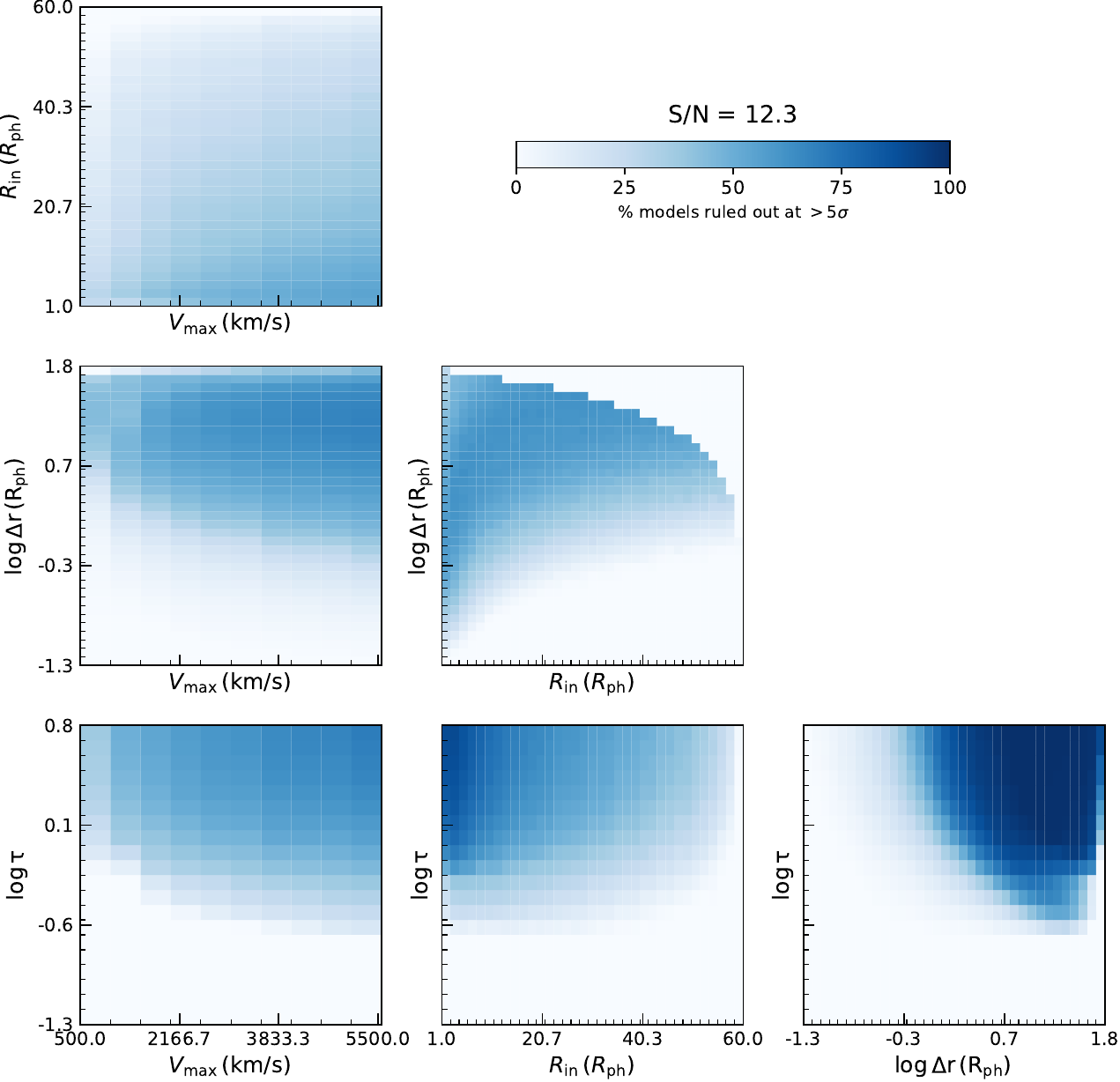}} \\
     \subcaptionbox{SN\,2021ek}{\includegraphics[width=0.3\textwidth]{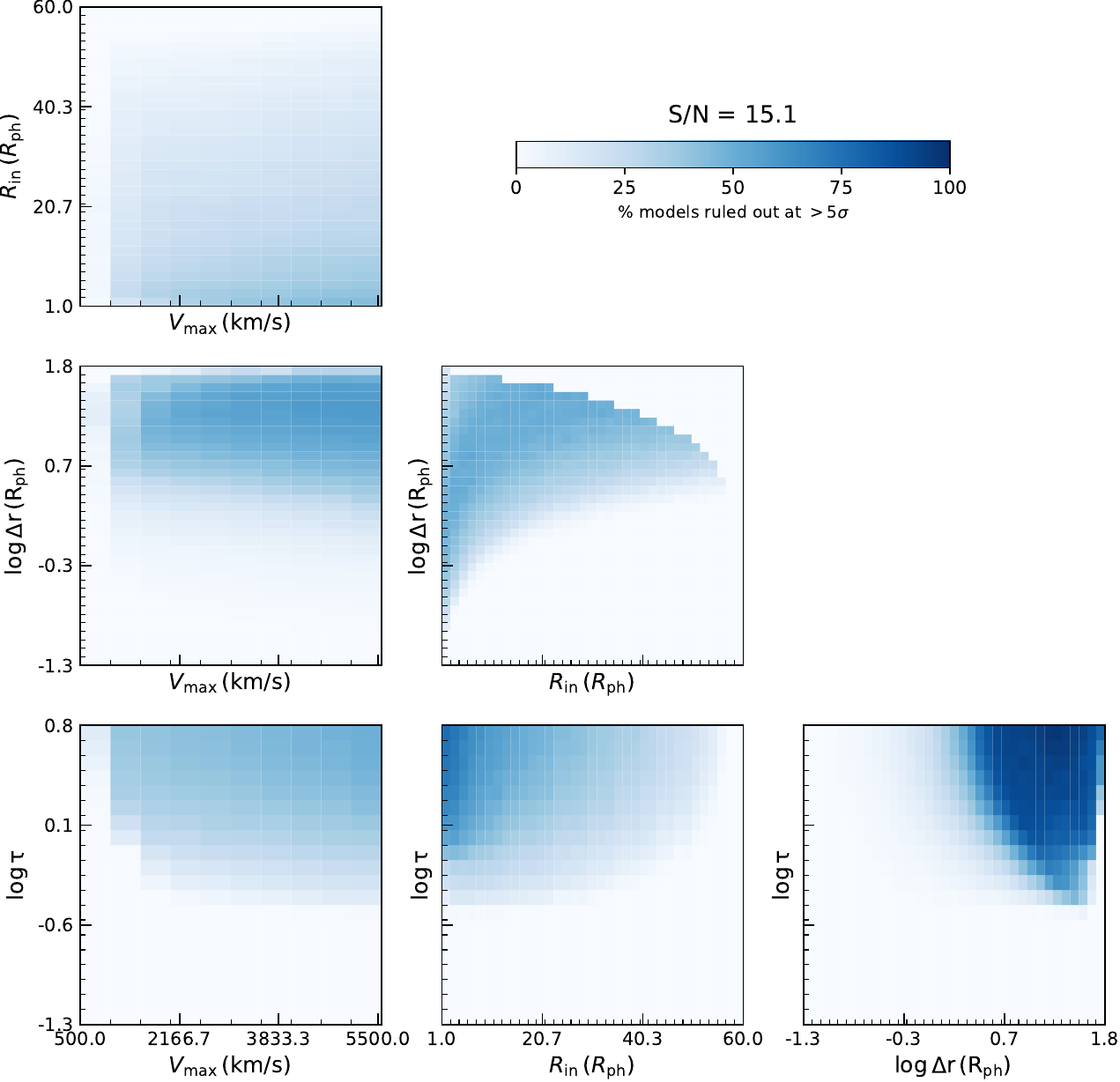}}
    \subcaptionbox{SN\,2021hpx}{\includegraphics[width=0.3\textwidth]{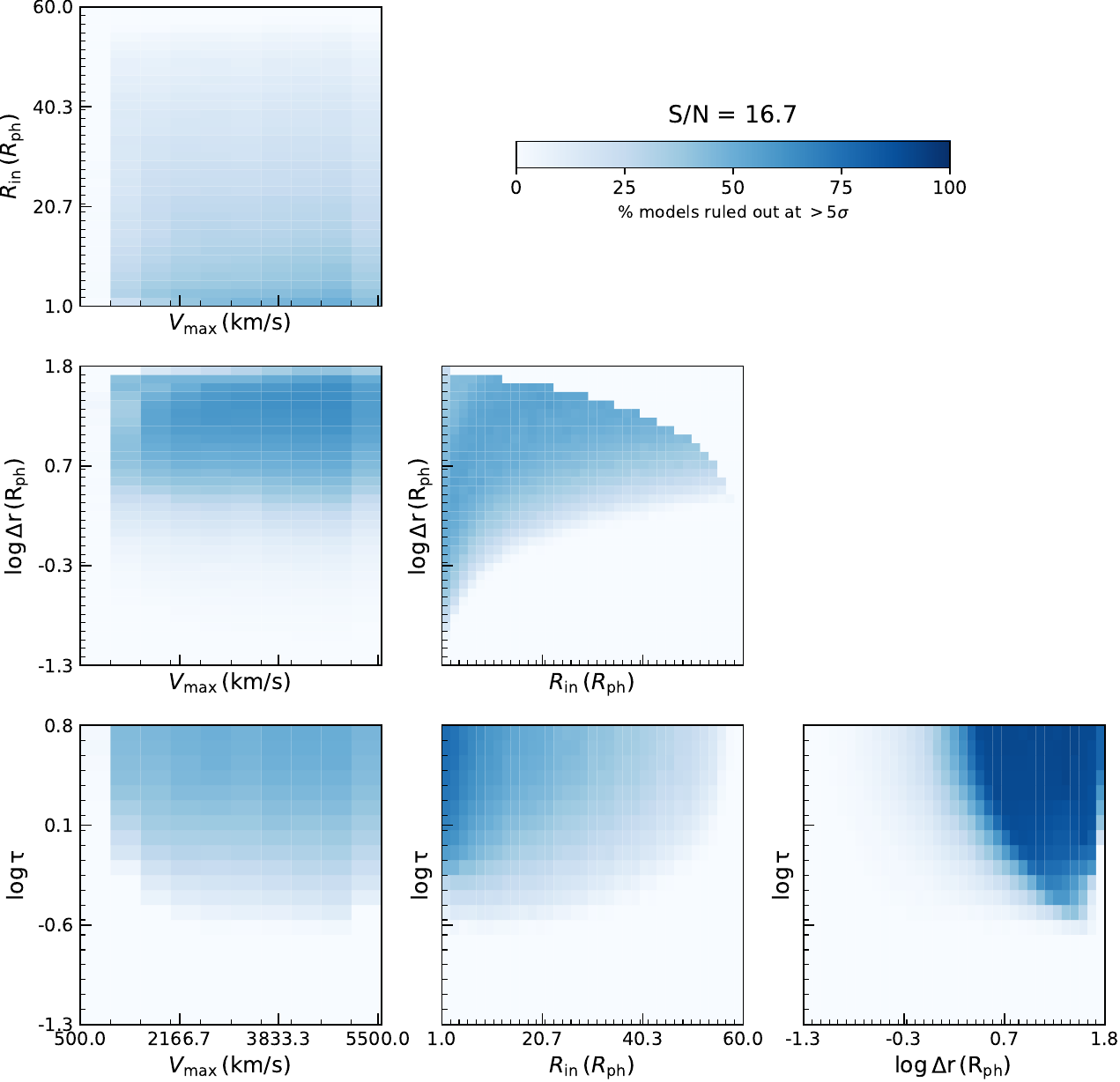}}
     \subcaptionbox{SN\,2020rmv}{\includegraphics[width=0.3\textwidth]{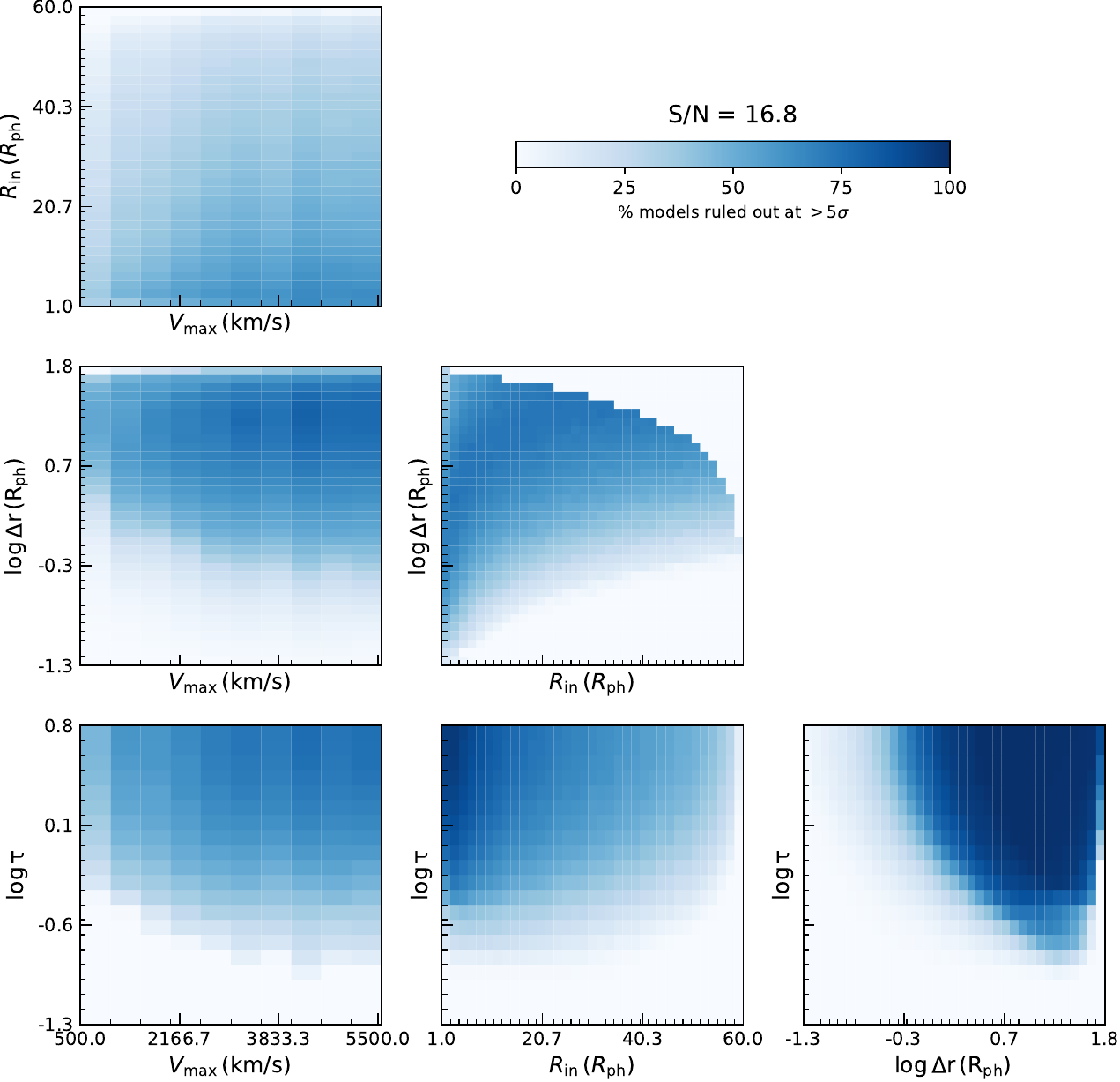}} \\

    \caption{\textit{Continued:} Corner plots showing, in a colorbar, the percentage of CSM models ruled out for nine out of the 16 events in our sample without CSM detections, shown as a function of an increasing S/N. The parameters correspond to the maximum velocity $v_{\rm max}$, inner radius $R_{\rm in}$, thickness $\Delta r$ and optical depth $\tau$.} 
    \label{fig:non_dete_cp_app}
\end{figure*}

\begin{figure*}[h]
  \centering
     \subcaptionbox{iPTF13ajg}{\includegraphics[width=0.3\textwidth]{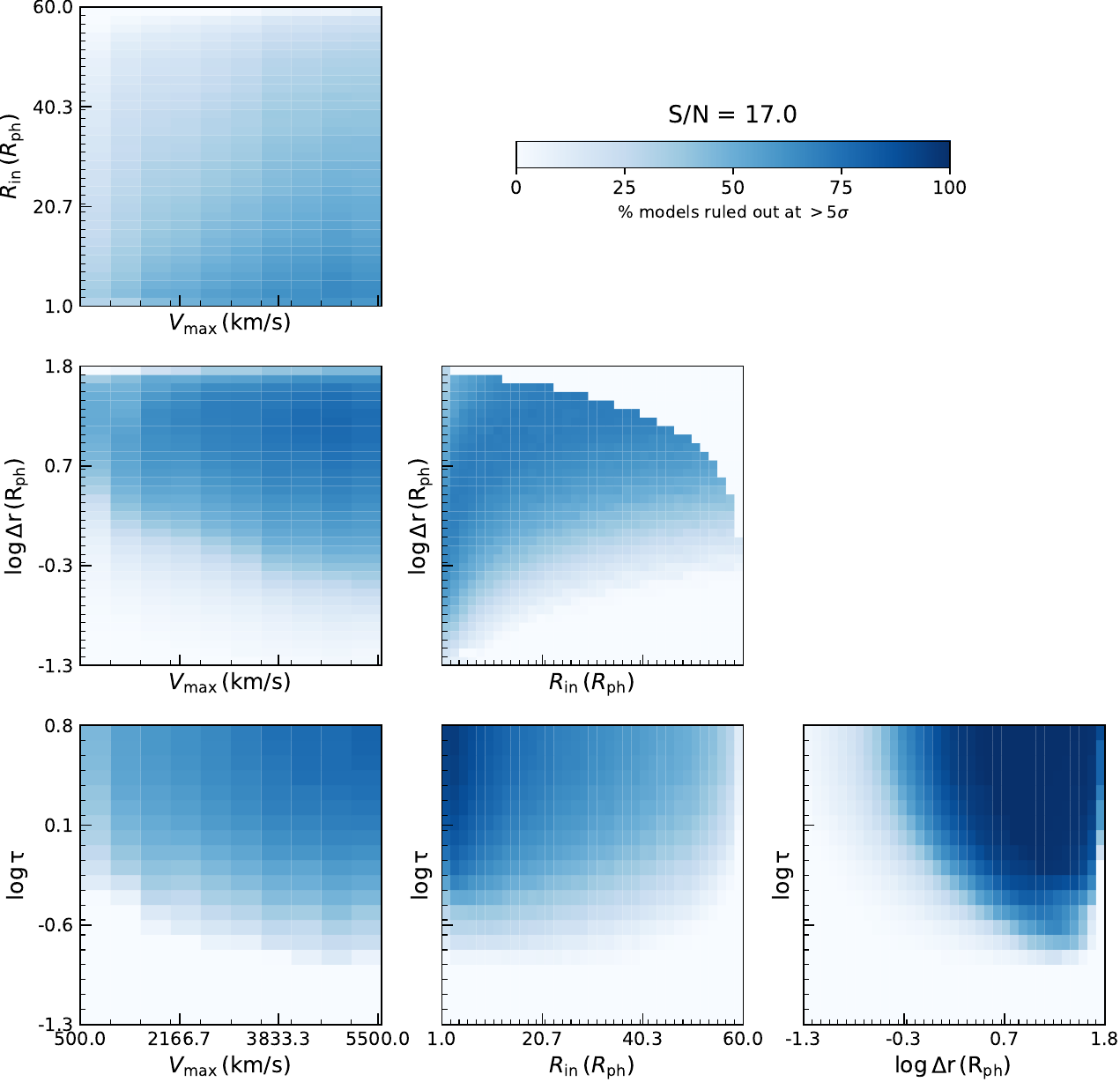}}
    \subcaptionbox{OGLE15qz}{\includegraphics[width=0.3\textwidth]{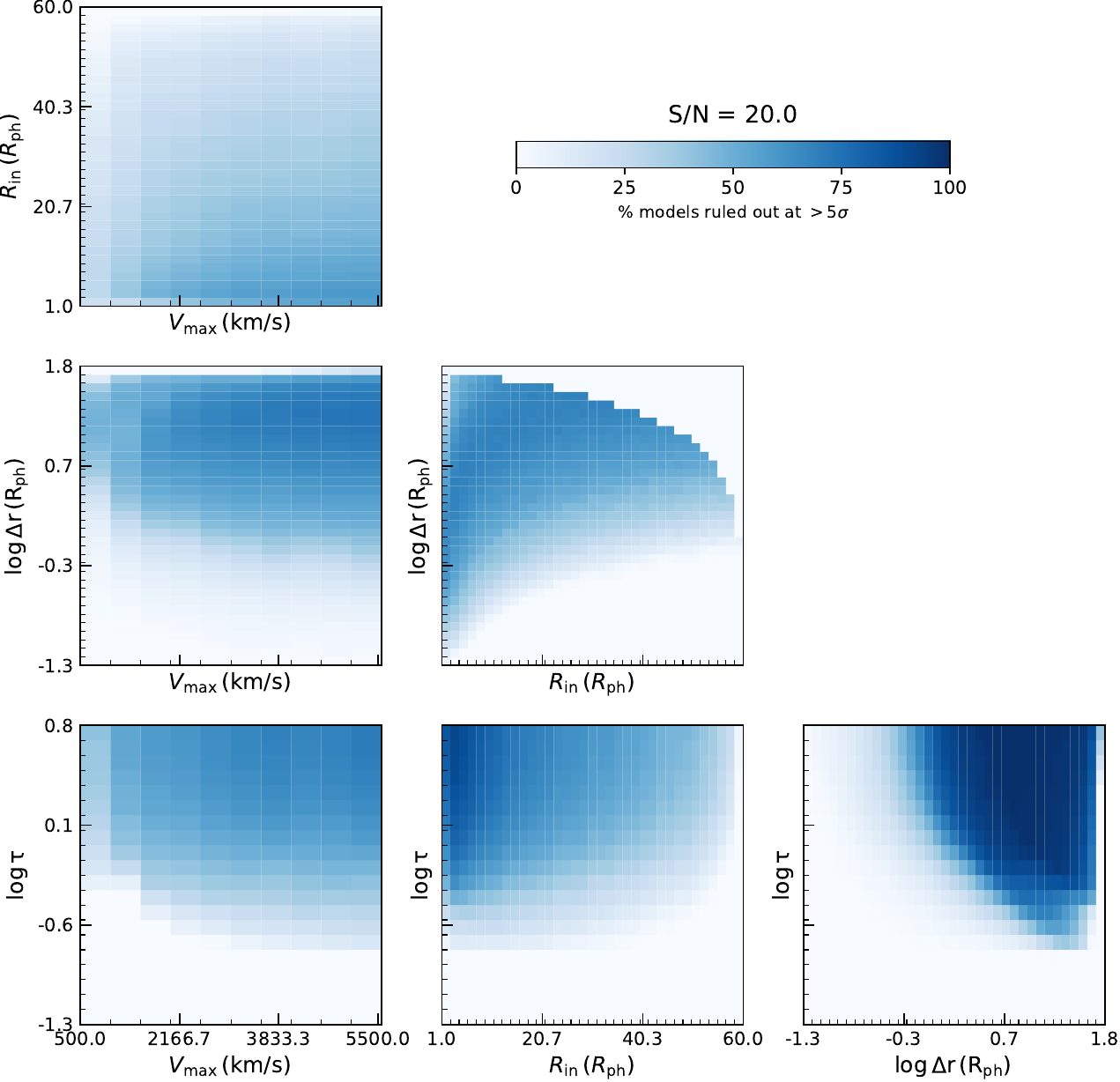}}
    \subcaptionbox{SN2020zbf}{\includegraphics[width=0.3\textwidth]{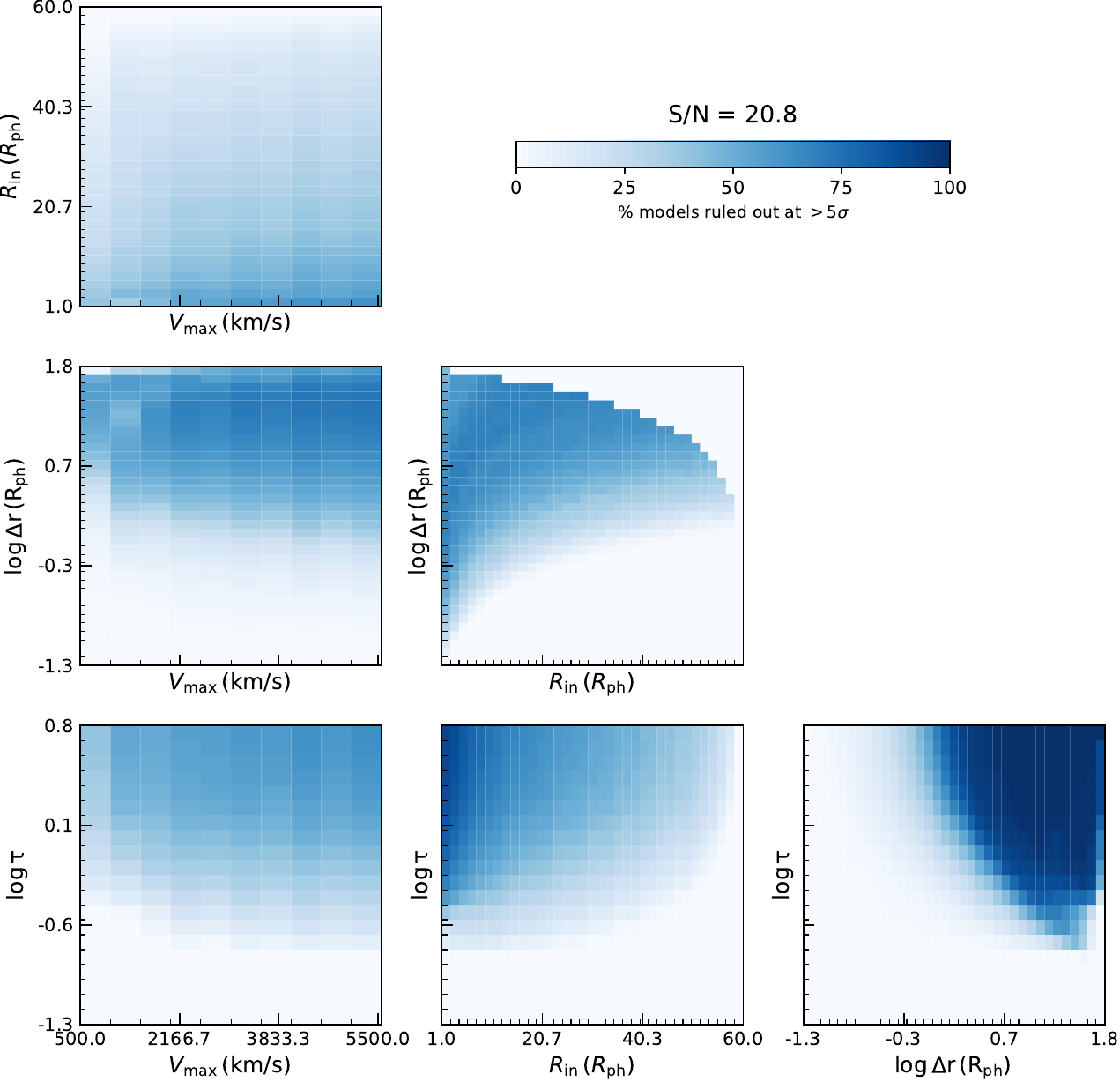}}\\
\caption{Continued: Corner plots showing, in a colorbar, the percentage of CSM models ruled out for three out of the 16  events in our sample without CSM detections, shown as a function of an increasing S/N. The parameters correspond to the maximum velocity $v_{\rm max}$, inner radius $R_{\rm in}$, thickness $\Delta r$ and optical depth $\tau$.}
    \label{fig:non_dete_cp_app_2}
\end{figure*}

\end{appendix}

\end{document}